\documentclass[fleqn,usenatbib]{mnras}
\usepackage{newtxtext,newtxmath}
\usepackage[modulo,switch]{lineno}
\usepackage{amsmath}
\usepackage{xcolor}
\usepackage[T1]{fontenc}
\usepackage{graphicx}

\DeclareRobustCommand{\VAN}[3]{#2}
\let\VANthebibliography\thebibliography
\def\thebibliography{\DeclareRobustCommand{\VAN}[3]{##3}\VANthebibliography}

\newcommand{\gsim}{\lower.5ex\hbox{$\; \buildrel > \over \sim \;$}}
\newcommand{\lsim}{\lower.5ex\hbox{$\; \buildrel < \over \sim \;$}}

\newcommand{\kms}{\ifmmode {\rm km\,s}^{-1} \else km\,s$^{-1}$ \fi}

\newcommand{\ergcms}{\ifmmode {\rm ergs\,cm}^{-2}\,{\rm s}^{-1} \else ergs\,cm$^{-2}$\,s$^{-1}$\fi}
\newcommand{\ergcmsA}{\ifmmode{\rm ergs}\, {\rm cm}^{-2}\,{\rm s}^{-1}\,{\rm\AA}^{-1} \else ergs\, cm$^{-2}$\, s$^{-1}$\, \AA$^{-1}$\fi}
\newcommand{\ergcmsHz}{\ifmmode{\rm ergs\,cm}^{-2}\,{\rm s}^{-1}\,{\rm Hz}^{-1} \else ergs\,cm$^{-2}$\,s$^{-1}$\,Hz$^{-1}$\fi}
\newcommand{\phcms}{\ifmmode {\rm ph\,cm}^{-2}\,{\rm s}^{-1} \else ,ph\,cm$^{-2}$\,s$^{-1}$\fi}
\newcommand{\phcmsA}{\ifmmode {\rm ph\,cm}^{-2}\,{\rm s}^{-1}\,{\rm\AA}^{-1} \else ph\,cm$^{-2}$\,s$^{-1}$\,\AA$^{-1}$\fi}

\newcommand\Msun{\ifmmode M_{\odot} \else $M_{\odot}$\fi}
\newcommand\msun{\ifmmode M_{\odot} \else $M_{\odot}$\fi}
\newcommand\Lsun{\ifmmode L_{\odot} \else $L_{\odot}$\fi}
\newcommand\Zsun{\ifmmode Z_{\odot} \else $Z_{\odot}$\fi}
\newcommand\mpyr{\ifmmode \Msun\,{\rm yr}^{-1} \else $\Msun\,{\rm yr}^{-1}$ \fi}

\newcommand{\Luv}{\ifmmode L_{1450} \else $L_{1450}$\fi}
\newcommand{\Lop}{\ifmmode L_{5100} \else $L_{5100}$\fi}
\newcommand{\Lthree}{\ifmmode L_{3000} \else $L_{3000}$\fi}
\newcommand{\lledd}{\ifmmode L/L_{\rm Edd} \else $L/L_{\rm Edd}$\fi}
\newcommand{\ledd}{\ifmmode L_{\rm Edd} \else $L_{\rm Edd}$\fi}
\newcommand{\lamLlam}{\ifmmode \lambda L_{\lambda} \else $\lambda L_{\lambda}$\fi}
\newcommand{\lbol} {\ifmmode L_{\rm bol} \else $L_{\rm bol}$\fi}
\newcommand{\llbol}{\ifmmode \log\left(\lbol/\ergs\right) \else $\log\left(\lbol/\ergs\right)$\fi}

\newcommand{\fuv}{\ifmmode f_{\lambda}\left(1450\AA\right) \else $f_{\lambda}\left(1450 {\rm \AA}\right)$\fi}
\newcommand{\fthree}{\ifmmode f_{\lambda}\left(3000\AA\right) \else $f_{\lambda}\left(3000{\rm \AA}\right)$\fi}
\newcommand{\fH}{\ifmmode f_{\lambda}\left(1.65\micron\right) \else
$f_{\lambda}\left(1.65\micron\right)$\fi}

\newcommand{\mbh}{\ifmmode M_{\rm BH} \else $M_{\rm BH}$\fi}
\newcommand{\lmbh}{\ifmmode \log\left(\mbh/\Msun\right) \else $\log\left(\mbh/\Msun\right)$\fi}

\def\dif{\mathop{}\hphantom{\mskip-\thinmuskip}\mathrm{d}}%
\let\daccent\d
\gdef\d{\ifmmode\dif\else\expandafter\daccent\fi}

\title[Symmetry on SFMS]{Symmetry in Fundamental Parameters of Galaxies on the Star-forming Main Sequence}

\author[Zhicheng He et al.]{
Zhicheng He,$^{1,2}$\thanks{E-mail: zcho@ustc.edu.cn}
Enci Wang,$^{1,2}$\thanks{E-mail: ecwang16@ustc.edu.cn}
Luis C. Ho,$^{3}$
Huiyuan Wang,$^{1,2}$
Yong Shi,$^{4}$
Xu Kong,$^{1,2}$
Tinggui Wang$^{1,2}$
\\
$^{1}$Department of Astronomy, University of Science and Technology of China, Hefei 230026, China\\
$^{2}$School of Astronomy and Space Science, University of Science and Technology of China, Hefei, Anhui 230026, China\\
$^{3}$Kavli Institute for Astronomy and Astrophysics, Peking University, Beijing 100871, China\\
$^{4}$School of Astronomy and Space Science, Nanjing University, Nanjing 210023, China\\
}

\date{Accepted XXX. Received YYY; in original form ZZZ}

\pubyear{2025}

\begin{document}
\maketitle
\begin{abstract}
 The Star-Forming Main Sequence (SFMS) serves as a critical framework for understanding galaxy evolution, highlighting the relationship between star formation rates (SFR) and stellar masses $M_*$ across cosmic time, for star forming galaxies. Despite its significance, the origin of the 0.3–0.4 dex dispersion in the SFMS remains a topic of debate. Uncovering the origin of dispersion is crucial for understanding the evolution of galaxies. Using a large sample of approximately 500,000 galaxies, we reveal a set of clear and systematic symmetries in the distribution of key structural properties—effective radius ($R_{\rm e}$), stellar surface density ($M_*/R_{\rm e}^2$), and morphology—on the SFMS. This symmetry implies that galaxies with high (above SFMS) and low (below SFMS) SFRs share similar fundamental parameters. Moreover, galaxies with smaller $R_{\rm e}$ or higher $M_/R_{\rm e}^2$ exhibit greater dispersion in SFR. This dispersion reflects the response to fluctuations in cosmic accretion flows, while the SFR itself represents the time-averaged effect over the gas consumption timescale. Shorter gas consumption timescales, associated with higher $M_/R_{\rm e}^2$, lead to greater SFR dispersion. Our results reveal that the variation of SFR originates from the oscillation of accretion flow and is regulated by the stellar surface density.
\end{abstract}


\section{Introduction}
The star formation main sequence (SFMS), e.g.,~ \cite{brinchmann2004,daddi2007,elbaz2007,elbaz2011,noeske2007,schiminovich2007,franx2008,peng2010,rodighiero2010,whitaker2012,whitaker2014,speagle2014,renzini2015,schreiber2015}, which describes the tight correlation between a galaxy's stellar mass ($M_*$) and its star formation rate (SFR), is a cornerstone in understanding galaxy evolution. The SFMS provides a crucial framework for interpreting galaxy growth, as its shape and evolution reflect the physical processes driving star formation across cosmic time. However, beyond the global trends, the intrinsic scatter or dispersion of SFMS has emerged as a topic of considerable interest. 
Many observational studies reveal that the SFMS scatter depends on stellar mass, showing a U-shaped $\sigma_{\rm SFR}$-$M^*$ relation, with increased dispersion at both the low- and high-mass ends (e.g.,~\citealt{guo2015,kurczynski2016,santini2017,boogaard2018,davies2019,katsianis2019,huang2023,clarke2024,davies2025a,davies2025b}). This dispersion is not merely a statistical artifact, but encodes critical information about the diverse processes affecting galaxies. In general, the mechanisms that affect the dispersion of SFMS as well as SFR can be roughly divided into two categories: internal factors, such as galaxy mass, size, morphology, gas content and star formation efficiency (e.g.,~\citealt{genzel2015, tacconi2018, Yesuf2021, Yu2021, Yu2022, Yu2022b}), usually, these factors are also interrelated,
feedback processes (e.g.,~\citealt{fabian2012,ishibashi2012,zubovas2012,cicone2014,king2015,he2019, he2022,chen2022,veilleux2020,he2024}), including those induced by active galactic nuclei (AGN) and stellar winds; external influences, such as dense environments or galaxy mergers (e.g.,~\citealt{peng2010,rasmussen2012,ayromlou2021,old2020}).
Meanwhile, cosmic evolution and stochastic gas accretion histories can also be considered to provide contributions to this dispersion
(e.g.,~\citealt{speagle2014,wang2019,Caplar-19, wang2020a,tacchella2020, popesso2023}). However, which of the above factors dominates or how they coordinate and interact with each other remains a highly contentious and unresolved mystery in understanding the SFMS dispersion.

The mass and size are two fundamental parameters of galaxies that shape their gravitational fields. The efficiency for converting gas into stars (SFE, defined as the ratio of star formation rate to gas mass) is found to be strongly correlated to the stellar mass surface density ($\Sigma_*$) of galaxies, i.e. SFE $\propto \Sigma_* ^{0.5}$ \citep{shi2011,shi2018}. This empirical law shows the role of existing stars in controlling the SFE, which can be interpreted in terms of the freefall time of cold gas collapse in a stellar-dominated potential \citep{Dopita1994}.
The $\Sigma_*$ is by definition correlated with galaxy mass and size. Therefore, we first investigate the roles of total stellar mass ($M_*$) and effective radius ($R_{\rm e}$) in shaping the galaxy's evolution and their contributions to the enigmatic and contentious origins of SFMS dispersion.
Here we utilized a large sample of approximately 600,000 galaxies derived from Sloan Digital Sky Survey Data Release 7 (SDSS DR7) to investigate the mechanism of SFMS dispersion.

\section{The sample description} \label{subsec:sample}
We cross-matched the GALEX-SDSS-WISE Legacy Catalog \citep{salim2007,salim2016,salim2018} and UPenn PhotDec Catalogs \citep{chen2012,mendel2013,meert2015,meert2016} to obtain total stellar masses 
$M_*$ and star formation rates from the former, along with morphology and effective radius $R_{\rm e}$ from the latter. The GALEX-SDSS-WISE Legacy Catalog (GSWLC) provides physical properties—such as stellar masses, dust attenuations, and star formation rates—for approximately 700,000 galaxies with SDSS redshifts below 0.3. The catalog integrates data from the Galaxy Evolution Explorer (GALEX), SDSS, and the Wide-field Infrared Survey Explorer (WISE). The UPenn PhotDec Catalog, created by researchers at the University of Pennsylvania, is a photometric classification catalog providing galaxy morphological classifications based on photometric data, primarily derived from SDSS imaging. This cross-match resulted in a sample of 561,433 galaxies, which was further refined by matching it with the Group Catalog \citep{yang2007} to classify central and satellite galaxies. The final sample consists of 490,123 galaxies, including 373,282 central galaxies and 116,841 satellite galaxies, primarily with $z < 0.2$. 
The effective radius we adopt is $r_\mathrm{tot}$ from the UPenn PhotDec Catalog FITS files, defined as the best-fitting model among the Dev, Ser, DevExp, and SerExp profiles. We then take the average of $r_\mathrm{tot}$ measured in the $r$, $g$, and $i$ bands as the effective radius of each galaxy.
We use the average value of $n_{\mathrm{S\acute{e}rsic}}$ in $r$, $g$ and $i$ bands as a pointer to the morphology of the galaxy.
In the sample, the number of galaxies with $n_{\mathrm{S\acute{e}rsic}}>2$ is 237,514, which have a morphology more similar to elliptical galaxies, and the number of galaxies with $n_{\mathrm{S\acute{e}rsic}}<2$ is 252,609, which have a morphology more similar to disk galaxies.

\begin{figure*}
\centering
\includegraphics[width=1.0\textwidth]{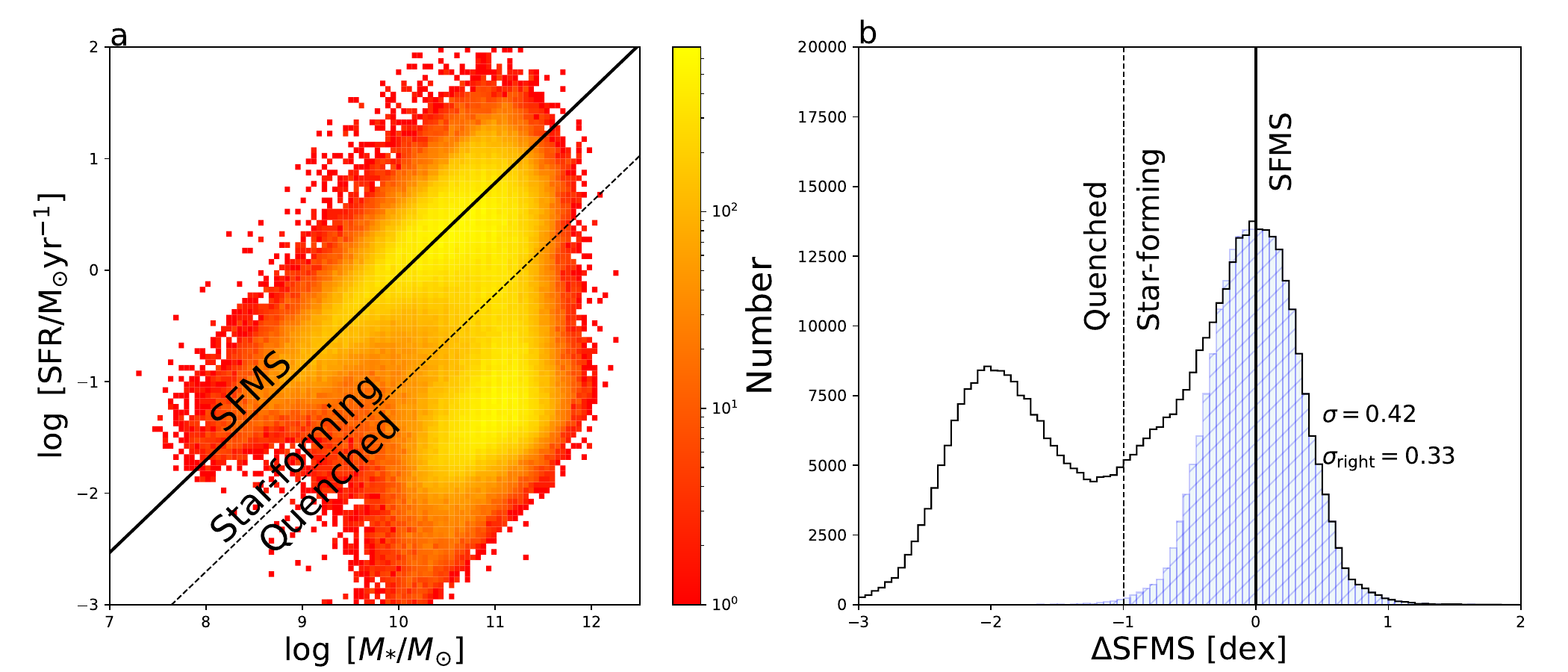}
\caption{\textbf{Star formation main sequence for the sample with 556,764 galaxies from
SDSS DR7.} (a), the black solid line represents the main sequence relationship of 
SFR and $M_*$: $\log {\rm SFR}=0.83\times \log {M_*}-8.33$. The dashed line represents the boundary between SF and quenched galaxies by shifting the SFMS down 1 dex.
(b) the deviation from the SFMS is defined as $\Delta \rm SFMS = \log \rm SFR - 0.83 \times \log M_{*} - 8.33$. The standard deviation of $\Delta \rm SFMS$ is approximately $\sigma = 0.42$ dex. The distribution of $\Delta \rm SFMS$ for SF galaxies is asymmetric, with an excess on the left, likely due to contributions from quenched or Green Valley galaxies. To correct this, the right-side distribution is symmetrically mirrored to the left, resulting in a more balanced distribution with a reduced dispersion of 0.33 dex.
}\label{fig1}
\end{figure*}

We present the $M_*$ versus SFR plot in panel a of  Fig. \ref{fig1}. To define the star formation main sequence (SFMS), we adopted the commonly used threshold of a specific star formation rate (sSFR) $10^{-11}\,\mathrm{yr^{-1}}$ as the initial boundary between SF and quenched galaxies (e.g.,~\citealt{wetzel2012,tamburri2014,citro2016,quai2018,donnari2021}). We then fitted the SFMS and excluded data points that deviated from the best-fit relation by more than 0.5\,dex before performing the next iteration. This procedure was repeated until the change in slope between two consecutive fits was less than 0.01.
The result SFMS is $\log {\rm SFR}=0.83\times \log {M_*}-8.33$ (solid line). Then, we move the SFMS down 1.0 dex as the boundary (dashed line) between SF galaxy and Quenched galaxy. According to this boundary, the number of SF galaxies and QG galaxies is 292,710 and 197,413 respectively. The SFMS typically shows a dispersion with 0.3-0.4 dex \citep{brinchmann2004, daddi2007,noeske2007,elbaz2011,wang2018,wang2019}. As shown in panel b of  Fig. \ref{fig1}, the dispersion (standard deviation) of SFMS is about $\sigma = 0.42$ dex. But it is obvious that the $\Delta \rm SFMS$ distribution of SF is asymmetric, with excess components on the left, possibly from contributions from QG or Green Valley galaxies. To address this, we focus only on the distribution on the right side of the SFMS and map it symmetrically to the left side, thereby obtaining a more balanced distribution. Under the condition, the dispersion of the SFMS
is 0.33 dex. Therefore, overall, the dispersion of SFMS is roughly 0.3-0.4 dex.

\begin{figure}
\centering
\includegraphics[width=0.45\textwidth]{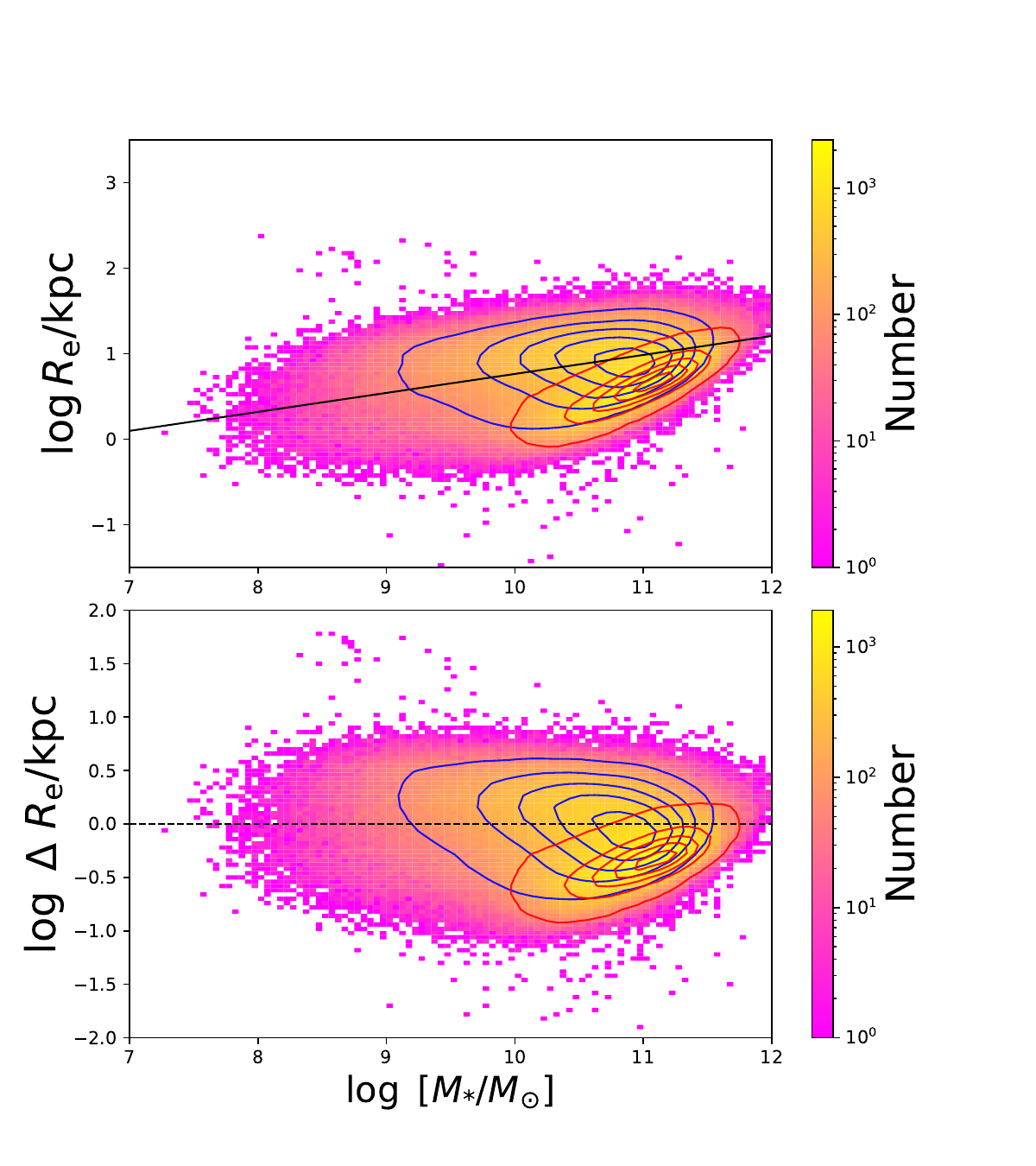}
\caption{\textbf{The regression relationship of $R_{\rm e}$ and $M_*$.} 
The best-fit regression between $R_{\rm e}$ and $M_*$ is given by $\log [R_{\rm e}/ \rm kpc] = 0.22 \times \log [M_*/ \Msun] - 1.46$. The residual of $R_{\rm e}$ is defined as 
$\Delta \log R_{\rm e} = \log R_{\rm e} - \log R_{\rm e}(M_*)$. Blue contours represent the isodensity lines for SF galaxies, while red contours represent those for quenched galaxies.
Since our focus is on the dispersion of the SFMS, we only analyze the dependence of 
$R_{\rm e}$ on $M_*$ for SF galaxies.}\label{fig2}
\end{figure}

\begin{figure}
\centering
\includegraphics[width=0.45\textwidth]{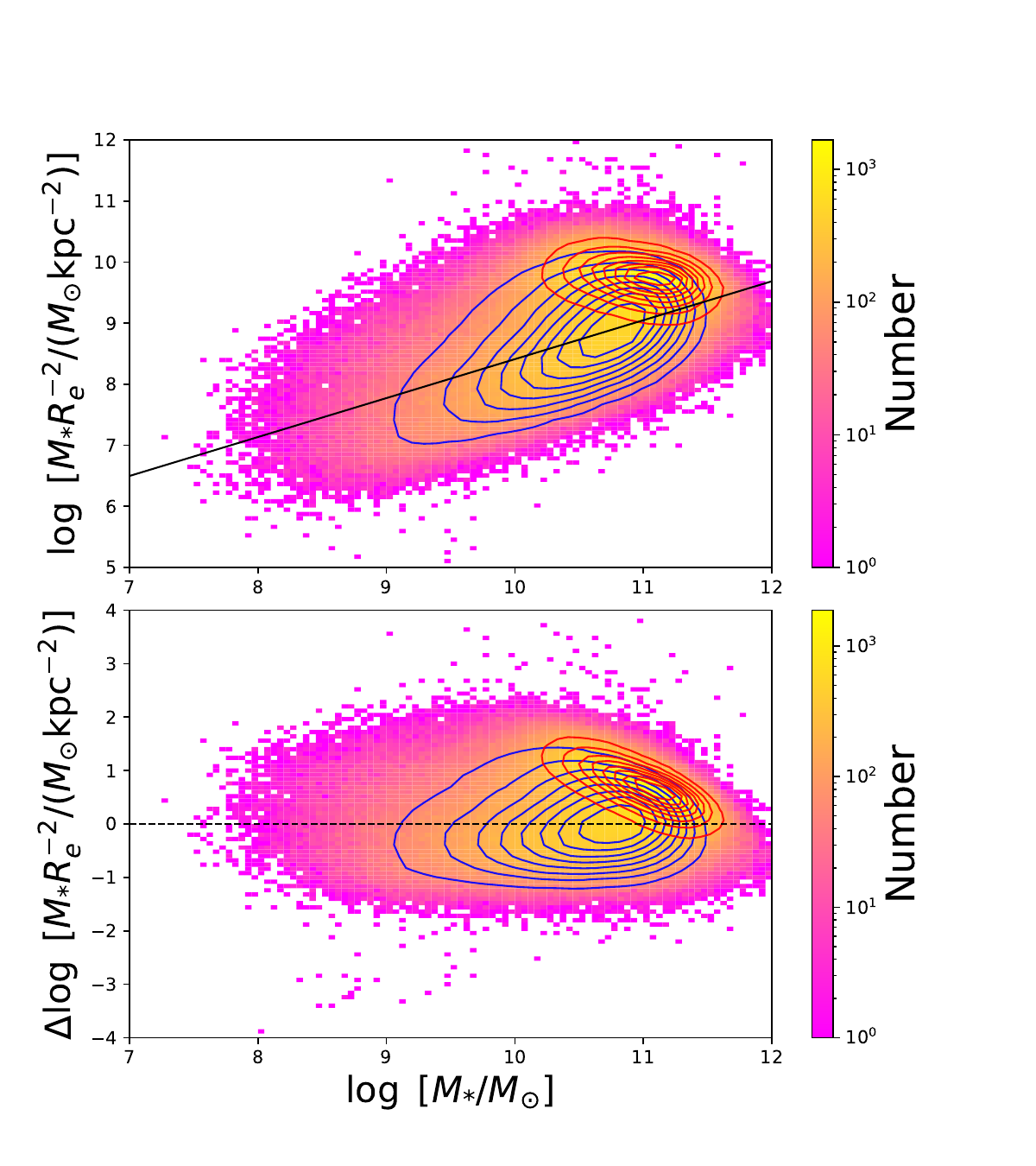}
\caption{\textbf{The regression relationship of $M_*/R_{\rm e}^2$ and $M_*$.} 
The best regression relationship for $M_*/R_{\rm e}^2$ and $M_*$ is $\log [M_*R_{\rm e}^{2-}/ \Msun \rm kpc^{-2}]$ = 0.66$\times \log [M_*/ \Msun]$ +1.80.
The residual of $M_*/R_{\rm e}^2$ are $\Delta \log [M_*/R_{\rm e}^2]$=$\log [M_*/R_{\rm e}^2]$ - $\log [M_*/R_{\rm e}^2](M_*)$. Blue contours represent the isodensity lines for SF galaxies, while red contours represent those for quenched galaxies.
Since our focus is on the dispersion of the SFMS, we only analyze the dependence of $M_*/R_{\rm e}^2$ on $M_*$ for SF galaxies.
}\label{fig3}
\end{figure}

The SFMS unfolds along the $M_*$. In this work, our focus is on highlighting the connection between the SFMS dispersion and $R_{\rm e}$ as well as the stellar surface density
$M_*/R_{\rm e}^2$. Therefore, in order to eliminate the influence of $M_*$, we must remove the dependence of $R_{\rm e}$ or $M_*/R_{\rm e}^2$ on $M_*$.
We remove their dependency on the $M_*$ via the linear regression model provided in the scikit-learn library \citep{pedregosa2011}.
As shown in panel a of  Fig. \ref{fig2}, the best regression relationship for $R_{\rm e}$ and $M_*$ is $\log [R_{\rm e}/ \rm kpc]$=0.22$\times \log [M_*/ \Msun]$ -1.46.
As shown in panel a of  Fig. \ref{fig3}, the best regression relationship for $M_*/R_{\rm e}^2$ and $M_*$ is $\log [M_*R_{\rm e}^{2-}/ \Msun \rm kpc^{-2}]$
= 0.66$\times \log [M_*/ \Msun]$ +1.80.
It should be pointed out that we are studying the problem of SFMS dispersion, so we only consider the dependence of $R_{\rm e}$ or $M_*/R_{\rm e}^2$ on $M_*$ in SF galaxies.
The residuals of $R_{\rm e}$ and $M_*/R_{\rm e}^2$ are $\Delta \log R_{\rm e}$=$\log R_{\rm e}$ - $\log R_{\rm e}(M_*)$ and $\Delta \log M_*/R_{\rm e}^2$=$\log M_*/R_{\rm e}^2$ - $\log M_*/R_{\rm e}^2(M_*)$, respectively (see bottom panels of  Fig. \ref{fig2} and \ref{fig3}).

\begin{figure}
\centering
\includegraphics[width=0.5\textwidth]{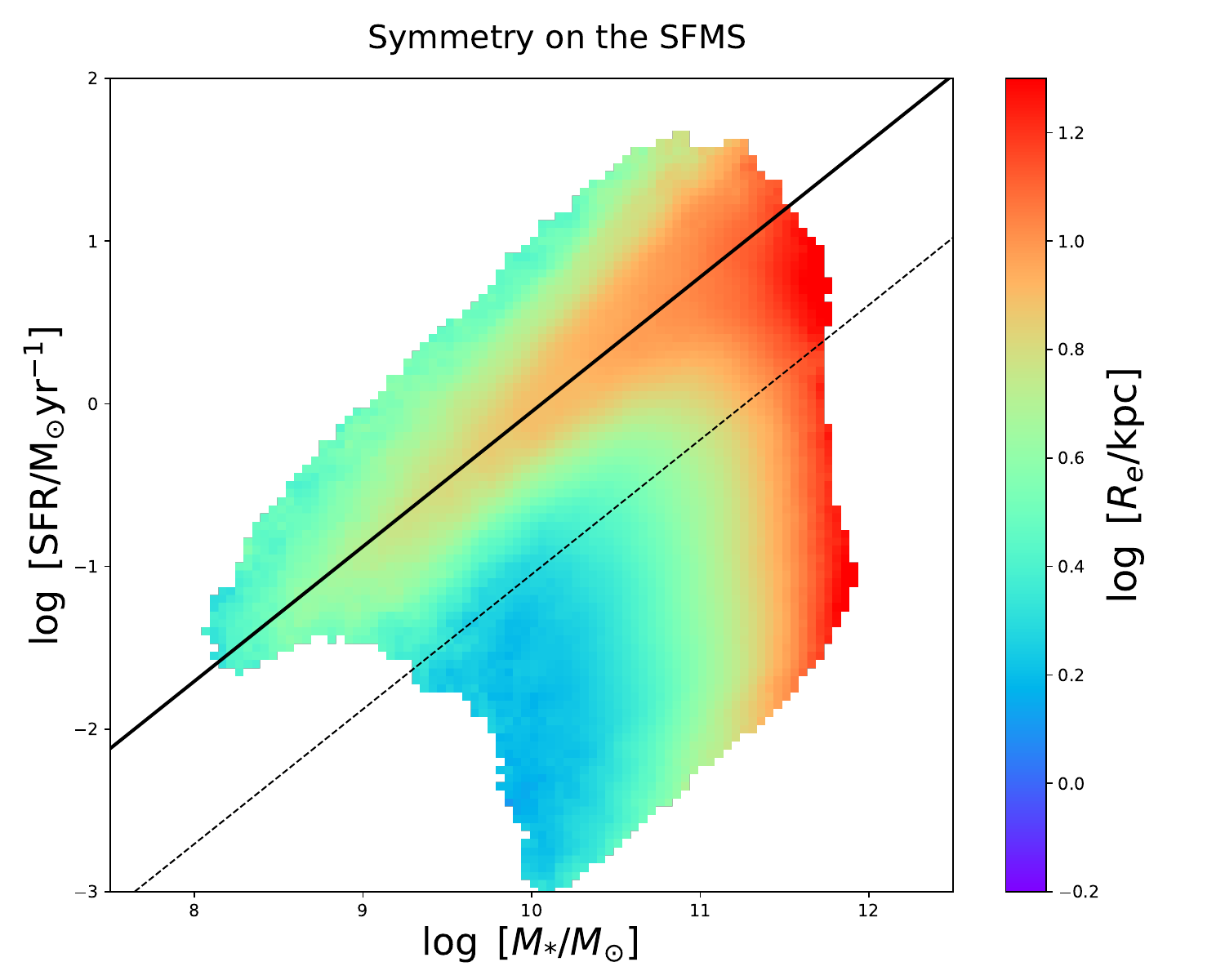}
\caption{\textbf{Symmetry for the galaxy effective radius on the SFMS.}
The galaxy effective radius ($R_{\rm e}$) exhibit symmetry along the SFMS, i.e., the closer to SFMS, the larger the $R_{\rm e}$.
The positive dependency between $R_{\rm e}$ and $M_*$ makes this symmetry appear less pronounced.}
\label{fig4}
\end{figure}

\begin{figure*}
\centering
\includegraphics[width=1.0\textwidth]{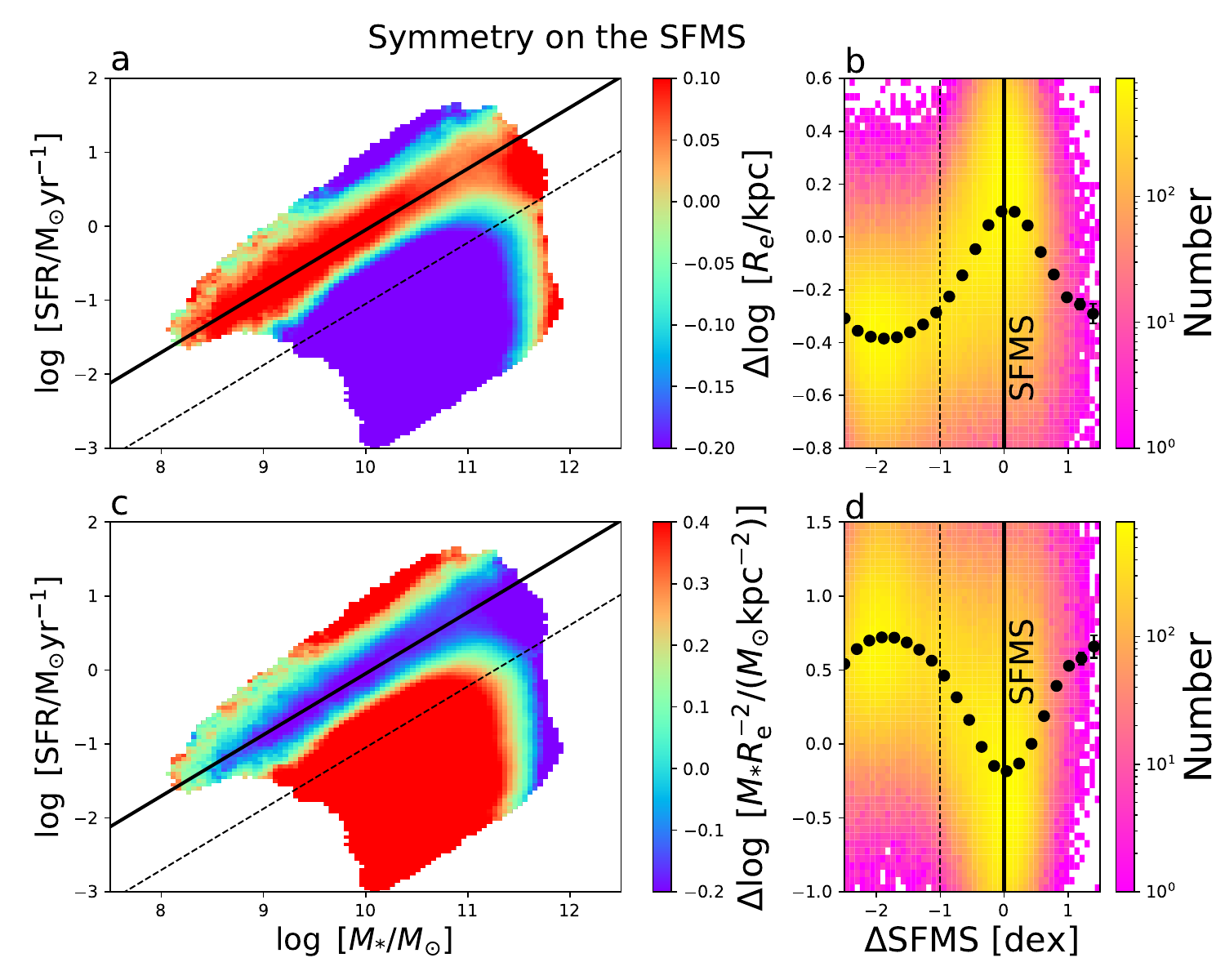}
\caption{\textbf{Symmetry Phenomenon on the SFMS.}
(a, c): After accounting for their dependence on $M_{}$, the residual galaxy effective radius ($R_{\rm e}$) and stellar surface density ($M_*/R_{\rm e}^2$) exhibit symmetry along the SFMS. Deviations correspond to smaller $R_{\rm e}$ or higher $M_*/R_{\rm e}^2$. To visualize this symmetry, the $R_{\rm e}$ and $M_*/R_{\rm e}^2$ values for each data point were averaged over a 0.1-dex surrounding range. The black straight line represents the SFMS, while the dashed line below marks the boundary between star-forming and quenched galaxies.
(b, d): On the SFMS, the average $\Delta \log R_{\rm e}$ reaches its maximum value and decreases as it deviates from the SFMS. Conversely, the average $\Delta \log M_*/R_{\rm e}^2$ is smallest on the SFMS and increases as it deviates from the SFMS.}
\label{fig5}
\end{figure*}

\begin{figure*}
\centering
\includegraphics[width=1.0\textwidth]{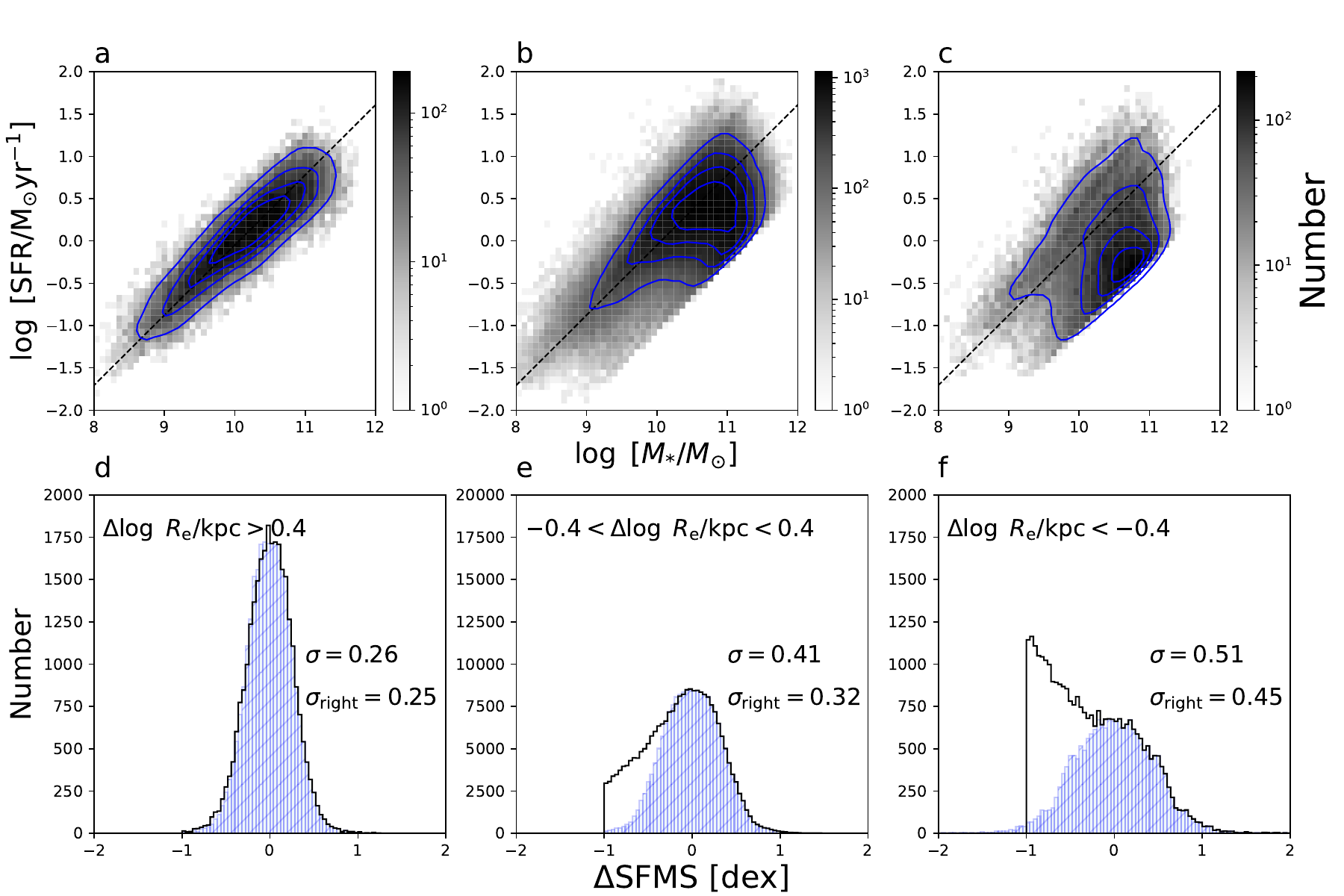}
\caption{\textbf{The SFMS dispersion increases as the galaxy effective radius decreases.}
(a), (b), (c): SFR versus $M_{*}$ plots for three residual radius intervals: $\Delta \log R_{\rm e}/\rm kpc > 0.4$, $-0.4 < \Delta \log R_{\rm e}/\rm kpc < 0.4$, and $\Delta \log R_{\rm e}/\rm kpc < -0.4$, respectively. Blue contours represent isodensity lines.
(d), (e), (f): SFMS dispersions for these intervals are 0.26, 0.41, and 0.51 dex, respectively, indicating an increase in dispersion as the residual radius decreases. The left-side asymmetry of the SFMS becomes more pronounced for smaller residual radii. To address this asymmetry, we symmetrically map the right-side distribution to the left, creating a more balanced distribution. Under these conditions, the SFMS dispersions are 0.25, 0.32, and 0.45 dex, respectively, still showing a trend of increasing dispersion with decreasing residual radius.
}\label{fig6}
\end{figure*}

\begin{figure*}
\centering
\includegraphics[width=1.0\textwidth]{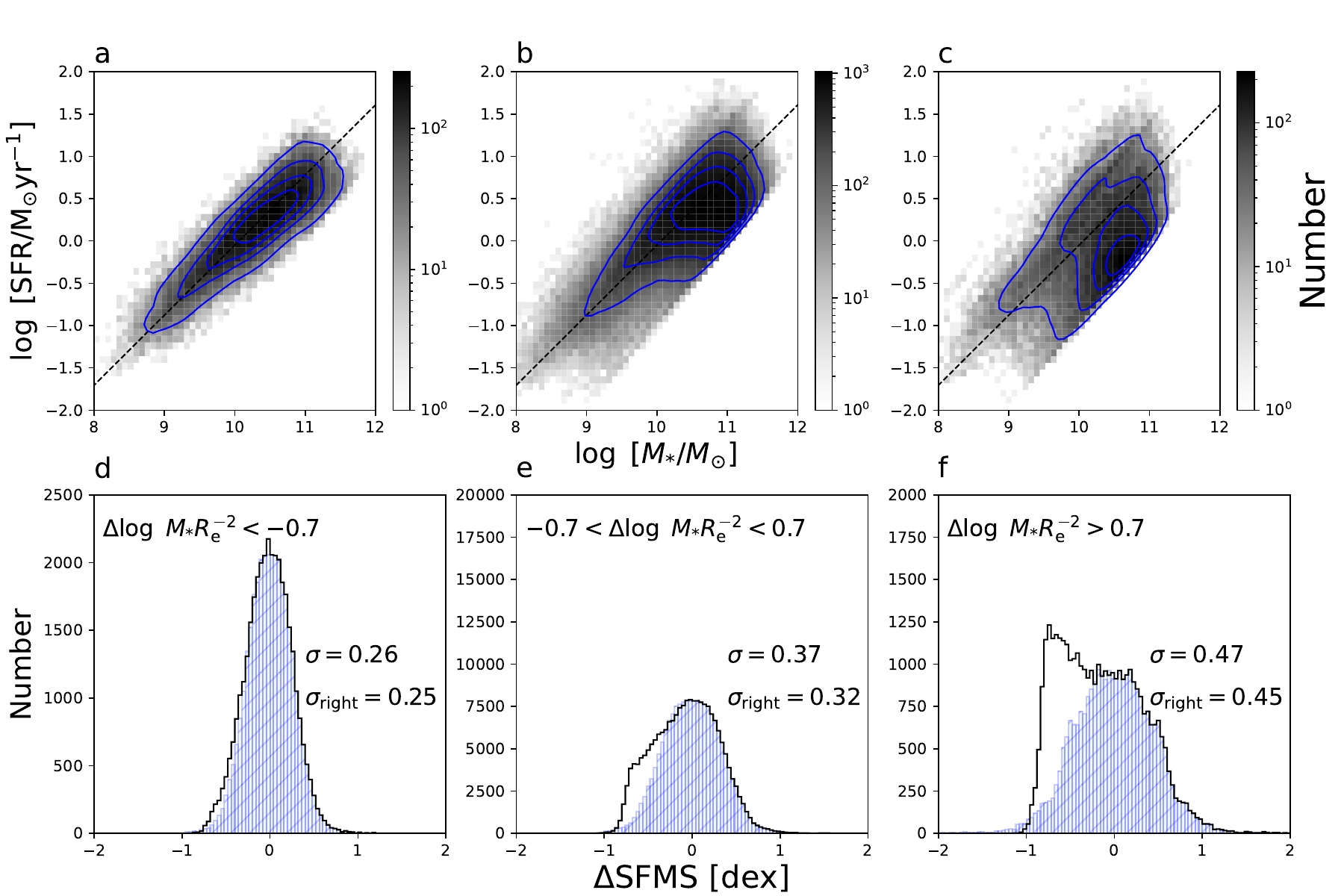}
\caption{\textbf{The SFMS dispersion increases as the stellar surface density increases.}
(a), (b), (c): SFR versus $M_{*}$ plots for three residual stellar surface density intervals: $\Delta \log M_*/R_{\rm e}^2 < -0.7$, $-0.7 < \Delta \log M_*/R_{\rm e}^2 < 0.7$, and $\Delta \log M_*/R_{\rm e}^2 > 0.7$, respectively. Blue contours represent isodensity lines.
(d), (e), (f): SFMS dispersions for these intervals are 0.26, 0.37, and 0.47 dex, respectively, indicating an increase in dispersion as the residual stellar surface density increases. The left-side asymmetry of the SFMS becomes more pronounced for larger residual stellar surface density. To address this asymmetry, we symmetrically map the right-side distribution to the left, creating a more balanced distribution. Under these conditions, the SFMS dispersions are 0.25, 0.32, and 0.45 dex, respectively, still showing a trend of increasing dispersion with increasing residual stellar surface density.
}\label{fig7}
\end{figure*}

\begin{figure*}
\centering
\includegraphics[width=1.0\textwidth]{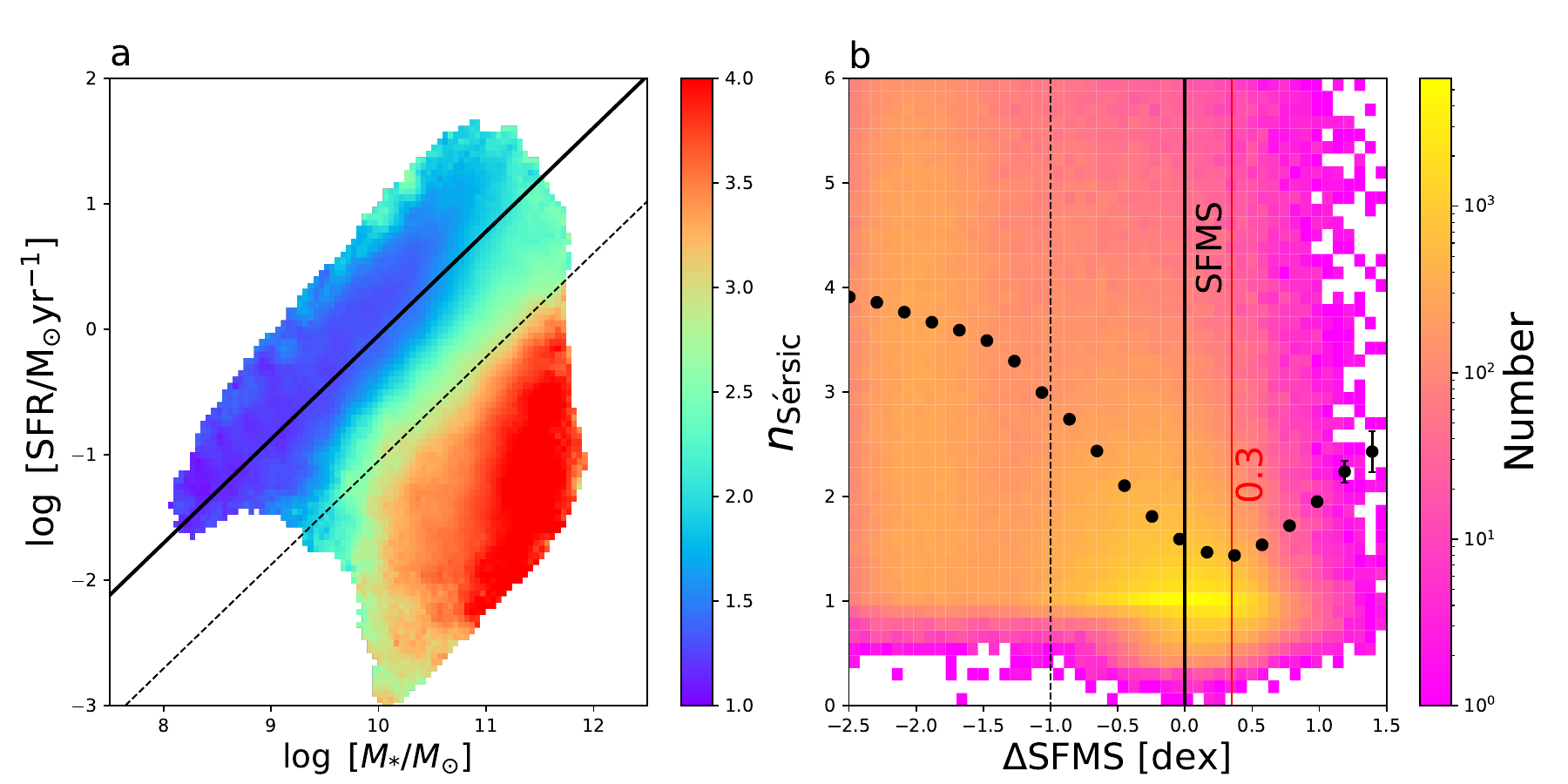}
\caption{\textbf{Symmetry Phenomenon for the indicator of galaxy morphology $n_{\mathrm{S\acute{e}rsic}}$ on the SFMS.}
a: The indicator of galaxy morphology $n_{\mathrm{S\acute{e}rsic}}$ exhibit symmetry along the SFMS. Deviations correspond to larger $n_{\mathrm{S\acute{e}rsic}}$. The black straight line represents the SFMS, while the dashed line below marks the boundary between star-forming and quenched galaxies. b: Slightly to the right of the main sequence , i.e., $\Delta \rm SFMS \simeq 0.3$, the average $n_{\mathrm{S\acute{e}rsic}}$ reaches its minimum value and increases as it deviates from the SFMS.}
\label{fig8}
\end{figure*}

\begin{figure}

\centering
\includegraphics[width=0.5\textwidth]{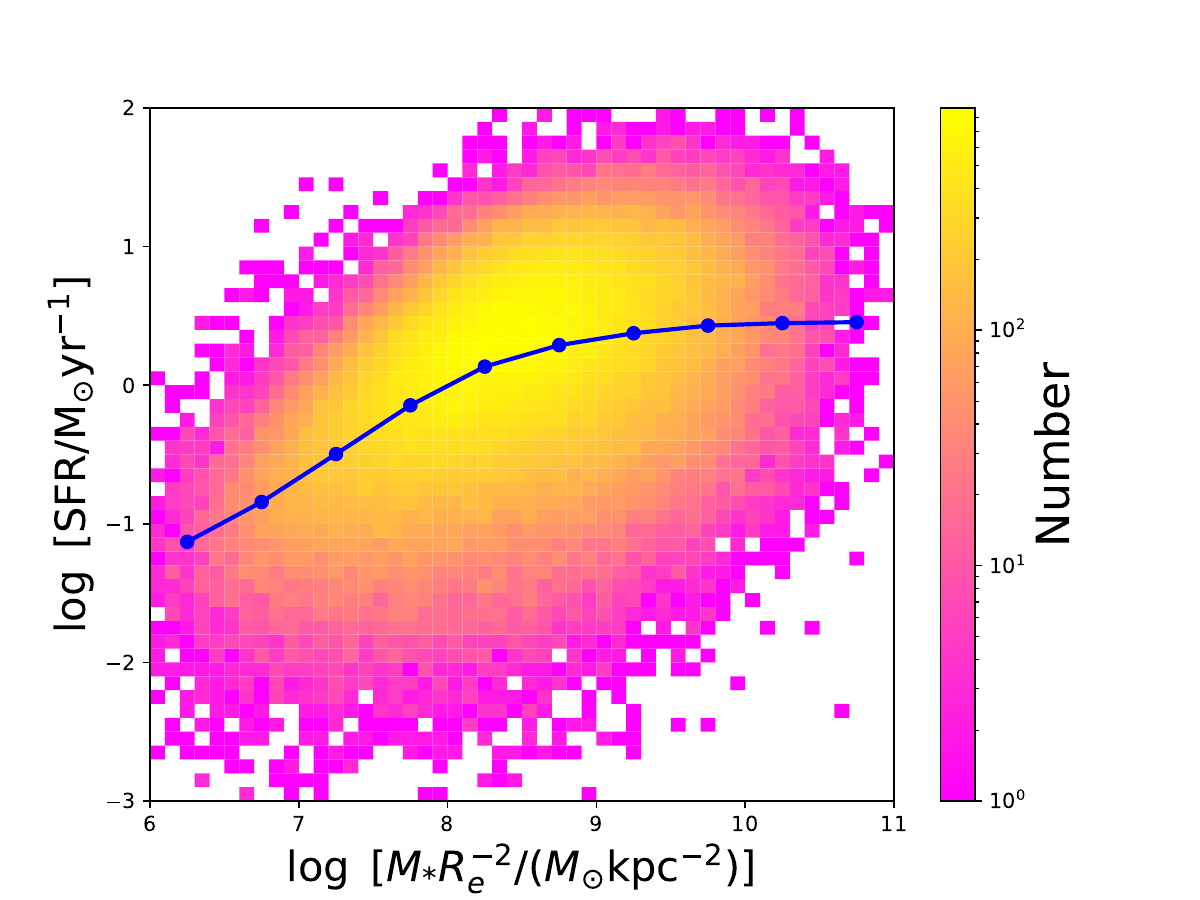}
\caption{\textbf{SFR as a function of the stellar surface density.
} SFR gradually increases with the stellar surface density.
The blue dots represent the average SFR at each bin of the stellar surface density.
}\label{fig9}
\end{figure}

\begin{figure}
\centering
\includegraphics[width=0.5\textwidth]{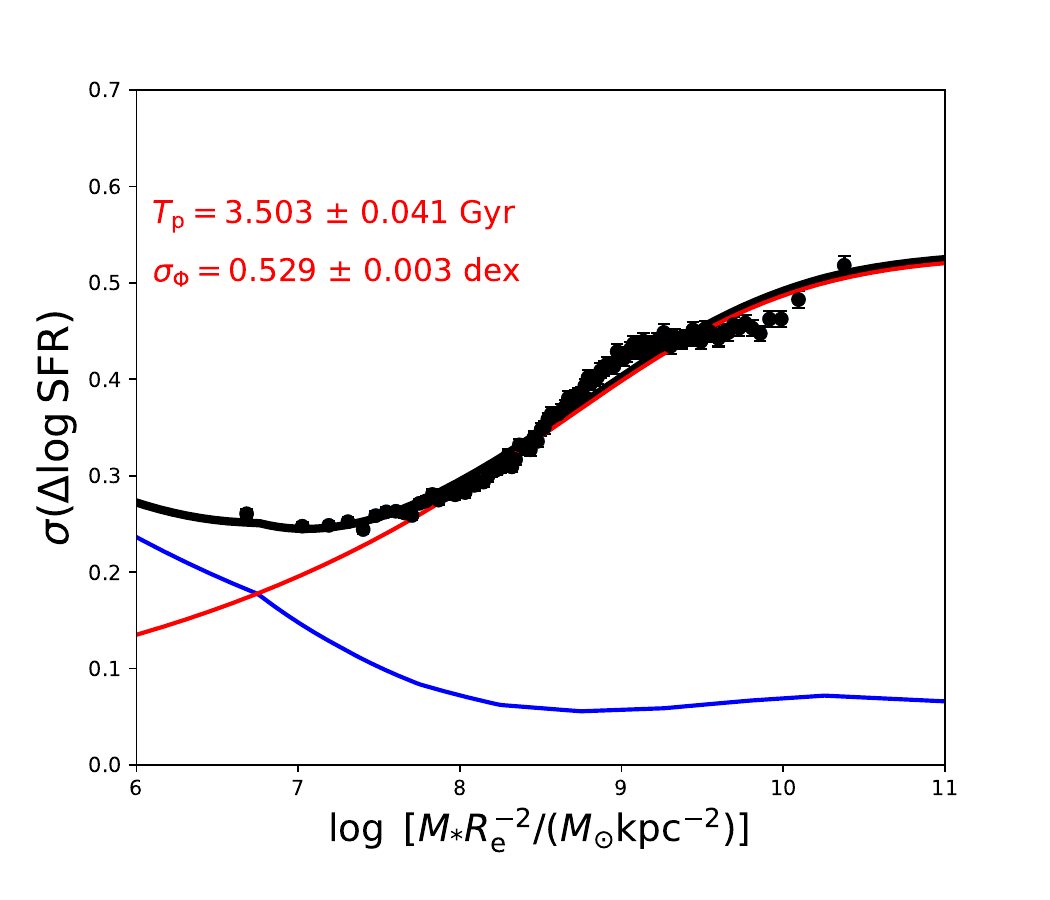}
\caption{\textbf{The SFR dispersion in SF galaxies.
} Overall, the SFR dispersion in SF galaxies increases with stellar surface
density. The blue curve represents the contribution
of statistical dispersion, while the red curve represents the contribution of the fluctuation of the galaxy gas accretion rate. The black curve is the best model fit and the corresponding fluctuation period of the gas accretion rate and the amplitude are listed.
}\label{fig10}
\end{figure}

\begin{figure}
\centering
\includegraphics[width=0.5\textwidth]{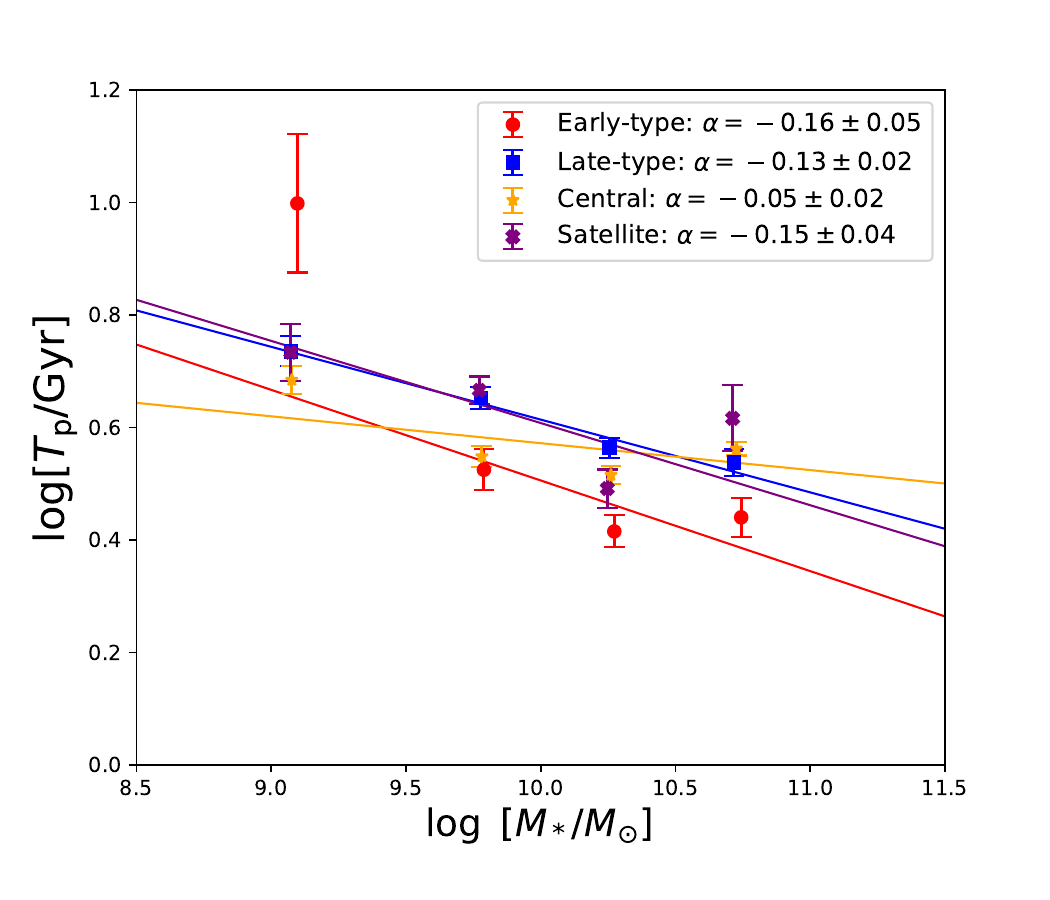}
\caption{\textbf{The gas accretion period decreases with the increase of galaxy mass.} 
The slope $\alpha$ to quantify the sensitivity of oscillation period $T_{\rm P}$ to galaxy mass $M_*$.
The decline of the accretion period with stellar mass is a bit steeper in early-type galaxies than in late types. 
The decline of the accretion period with stellar mass is steeper in satellites galaxies than in centrals types.
}\label{fig11}
\end{figure}

\section{Symmetry on the SFMS}
As shown in Fig.~\ref{fig4}, the effective radius of galaxies ($R_{\rm e}$) displays a symmetric trend around the SFMS: galaxies closer to the SFMS tend to have larger $R_{\rm e}$. 
Some previous literature has also reported similar signs of symmetry \citep{wuyts2011,tacchella2015,morselli2019}.
However, the positive correlation between $R_{\rm e}$ and stellar mass ($M_*$) makes this symmetry less evident.
In order to present this symmetry more clearly, we removed the dependency of $R_{\rm e}$ on $M_*$ and plotted the residual $\Delta \log R_{\rm e} $=$\log R_{\rm e}$- $\log R_{\rm e} (M_{*})$ (see  Fig. \ref{fig2}) on the $M_*$ and SFR plots. 
As shown in panel a of Fig.~\ref{fig5}, it is noteworthy that the residual effective radius, defined as $\Delta \log R_{\rm e}$, shows a more symmetric distribution than the $R_{\rm e}$ around the SFMS. Here, we define the deviation of the SFMS as $\Delta \rm SFMS $= $\log \rm SFR $ - 0.83 $\times \log M_{*}$-8.33.
As shown in panel b of Fig. \ref{fig5}, the residual $\Delta \log R_{\rm e}$ exhibits pronounced symmetry around $\Delta \rm SFMS = 0$. The average value of $\Delta \log R_{\rm e}$ is highest at $\Delta \rm SFMS = 0$ and decreases as it deviates from the SFMS. As shown in panels c and d of Fig.~\ref{fig5}, correspondingly, after removing the dependency on $M_*$, the residual stellar surface density $\Delta \log M_*/R_{\rm e}^2$ (see  Fig. \ref{fig3} for the dependence of stellar surface density on $M_*$) also shows noticeable symmetry, with deviations correlating to higher $\Delta \log M_*/R_{\rm e}^2$.

In Fig. \ref{fig6}, we divide the residual radius $\Delta \log R_{\rm e}$ into three intervals: $\Delta \log R_{\rm e} > 0.4$, $-0.4 < \Delta \log R_{\rm e} < 0.4$, and $\Delta \log R_{\rm e} < -0.4$ to assess whether the dispersion of the SFMS changes with the residual radius. The dispersion for these three intervals is 0.26, 0.41, and 0.51 dex, respectively, indicating an increasing trend in dispersion as the residual radius decreases. From the bottom panels of Fig. \ref{fig6}, it is evident that the asymmetry on the left side of the SFMS becomes more pronounced as the residual radius decreases. This is likely due to the gradual mixing of quenched galaxies.
To address this, we focus only on the distribution on the right side of the SFMS and map it symmetrically to the left side, thereby obtaining a more balanced distribution. Under these conditions, the dispersion of the SFMS is 0.25, 0.32, and 0.45 dex, respectively, still exhibiting the trend of increasing dispersion as the residual radius decreases.

We also investigated whether the dispersion of the SFMS varies with residual surface density. As shown in  Fig. \ref{fig7}, we divided the residual surface density into three intervals: $\Delta \log M_*/R_{\rm e}^2 < -0.7$, $-0.7 < \Delta \log M_*/R_{\rm e}^2 < 0.7$, and $\Delta \log M_*/R_{\rm e}^2 > 0.7$. Similar to the behavior of the residual radius, the dispersion in these intervals is 0.26, 0.37, and 0.47 dex, respectively, showing an increasing trend as the residual surface density increases. By again focusing only on the right side of the SFMS and mapping it to the left side symmetrically, we found the dispersion of the SFMS to be 0.25, 0.32, and 0.45 dex, respectively, maintaining the trend of increasing dispersion as the residual surface density increases.

Several studies \citep{Yesuf2021, Yu2021, Yu2022, Yu2022b} have established a connection between galaxy morphology and the SFMS. To further investigate this relationship, we analyzed the $\mathrm{S\acute{e}rsic}$ index, \( n_{\mathrm{S\acute{e}rsic}} \), as an indicator of galaxy morphology in the context of the SFMS.
As shown in panel a of  Fig.~\ref{fig8} , \( n_{\mathrm{S\acute{e}rsic}} \) exhibits a symmetric pattern around the SFMS, with deviations from the SFMS corresponding to higher values of \( n_{\mathrm{S\acute{e}rsic}} \). In panel (b), we observe that just to the right of the SFMS, around \( \Delta \rm SFMS \simeq 0.3 \), the average\( n_{\mathrm{S\acute{e}rsic}} \) reaches its minimum and increases as the deviation from the SFMS becomes larger. Our results indicate that when \( \Delta \rm SFMS \leq 0.3 \), \( n_{\mathrm{S\acute{e}rsic}} \) decreases as \( \Delta \rm SFMS \) increases. If a lower \( n_{\mathrm{S\acute{e}rsic}} \) corresponds to disk galaxies with more prominent spiral arms, then our findings align with previous studies, which suggest that galaxies with strong spiral structures are typically located above the SFMS within the range \( \Delta \rm SFMS \leq 0.3 \). However, our results also reveal that in the regime of the highest star formation rates, where \( \Delta \rm SFMS \geq 0.3 \), \( n_{\mathrm{S\acute{e}rsic}} \) increases with increasing \( \Delta \rm SFMS \). 
This may suggest that galaxies with moderately enhanced star formation (slightly higher than SFMS) are, on average, the most disk-dominated. Toward both lower and higher offsets, bulge growth 
(pre-quenching or starburst/merger-driven) increases the central concentration and thus raises the $\mathrm{S\acute{e}rsic}$ index.

In addition, we note that the SFMS is not necessarily well described by a single power-law over the entire stellar mass range. Several recent studies have reported a flattening or bending of the SFR--$M_\ast$ relation at the high-mass end (e.g., \citealt{tomczak2016,thorne2021,popesso2023}), suggesting that a broken power-law or smoothly bending functional form may better capture the observed behaviour of massive galaxies. As already visible in the left panel of Fig.~\ref{fig5}, our sample also exhibits a mild flattening toward the high-mass end. Adopting a functional form that explicitly accounts for this bending would slightly improve the symmetry of the residual distribution at high stellar masses. However, we find that reasonable variations in the MS definition---including allowing for high-mass bending---do not significantly alter the overall dispersion of the star-forming population nor our main conclusions regarding the symmetry of the SFMS scatter.

\section{Analysis of the Origin of Symmetry and Dispersion on SFMS}

The above results indicate that on both sides of SFMS, the effective radius, stellar surface density, and morphology of the galaxy are similar, and they all exhibit symmetry.
What is the causal relationship between the dispersion of SFMS and the symmetry of $R_{\rm e}$ or $M_*/R_{\rm e}^2$ ?

Let's start with the continuity equation for a galaxy first:
\begin{equation}
\frac{\mathrm{d} M_{\rm gas}(t)}{\mathrm{d} t}=
\Phi (t)- {\rm SFE} \cdot (1+\lambda)\cdot M_{\rm gas}(t),
\label{eq1}
\end{equation}
where $M_{\rm gas}$ is the mass of cold gas and $\lambda$
is the mass-loading factor of stellar wind \citep{lilly2013},
and $\Phi (t)$ is the gas inflow rate.
Based on the resolved $\Sigma_*$-sSFR-metallicity relation of a sample of MaNGA galaxies, the mass-loading factor was measured as:
$\lambda=\lambda_0(\Sigma_*/10^9[M_{\odot}{\rm kpc}^{-2}])^{-1/3}$ \citep{wang2019} with 
$\lambda_0\approx 1$.
The SFE is determined by the extended Schmidt law \citep{shi2011}:
\begin{equation}
\frac{\rm SFE}{\rm yr^{-1}}=10^{-10.28}
\left(\frac{\Sigma_*}{M_{\odot} \rm pc^{-2}}\right)^{0.48},
\label{eq2}
\end{equation}
where $\Sigma_*$ is the stellar mass surface density. For the galaxy as a whole, $\Sigma_*=M_*/(\pi R_{\rm e}^2)$.
The above differential Equation \ref{eq1} can be solved, yielding $M_{\rm gas}(t)$ in the form:
\begin{equation}
\begin{aligned}
M_{\rm gas}(t)&=&{\rm e}^{-t/\tau_{\rm dep}}\cdot
\left[{\rm Const} + \int_{-\infty}^{t} \Phi (t) \cdot {\rm e}^{t/\tau_{\rm dep}} \mathrm{d} t \right],\\
\\
&=&\tau_{\rm dep}\cdot
\frac{{\rm Const} + \int_{-\infty}^{t} \Phi (t) \cdot {\rm e}^{t/\tau_{\rm dep}} \mathrm{d} t }{\int_{-\infty}^{t}  {\rm e}^{t/\tau_{\rm dep}} \mathrm{d} t}
\label{eq3}
\end{aligned}
\end{equation}
where $\tau_{\rm dep}=\rm SFE^{-1}(1+\lambda)^{-1}$ is an
effective gas depletion timescale. $\rm Const$ is the initial phase. From the above solution of $M_{\rm gas}(t)$, we can see that $M_{\rm gas}(t)$ is determined by the sum of $\Phi (t)$ with ${\rm e}^{t/\tau_{\rm dep}}$ as the weight. We set the time, which is a certain period of time $\tau_{\rm dep}$ before time $t$, as $t_c=t-\tau_{\rm dep}$. Before time $t_c$, the weight is less than ${\rm e}^{t/\tau_{\rm dep}-1}$. Since ${\rm SFR}(t)=
{\rm SFE}\cdot M_{\rm gas}(t)$, SFR is mainly determined by the average value of $\Phi (t)$ weighted by ${\rm e}^{t/\tau_{\rm dep}}$ during the time from $t_c=t-\tau_{\rm dep}$ to $t$. This is equivalent to SFR being the result of smoothing $\Phi (t)$ over a time width of $\tau_{\rm dep}$.
Therefore, SFR has smaller fluctuations compared to $\Phi (t)$, and the specific relative fluctuations are determined by the relative sizes of $\tau_{\rm dep}$ and the fluctuation period $T_{\rm P}$ of $\Phi (t)$. Following \cite{wang2019}, the inflow rate is modeled as a sinusoidal function:
$\Phi (t)=\Phi (0)+{\rm A}\cdot \sin(2\pi/ T_{\rm P})$, where A is the amplitude of the fluctuation. \cite{wang2019} also derived the resulting fluctuation relation between SFR and $\Phi (t)$, expressed as:
\begin{equation}
\sigma_{\log \rm SFR}^{\Phi} =\frac{\sigma_{\log \Phi}}{[1+(2\pi \tau_{\rm dep}/T_{\rm P})^2]^{1/2}}.
\label{eq4}
\end{equation}
\cite{wang2019} also tested other forms of inflow functions $\Phi (t)$, such as inverse error function in logarithmic space [Inerf-$\log \Phi (t)$], which produces a Gaussian distribution of the inflow rate, but the resulting relationships of $\sigma_{\rm SFR}$ and $\sigma_{\Phi}$ were very similar (see Figures 12 and 13 in  \citealt{wang2019}). The most important thing here is the period and amplitude of fluctuations for $\Phi (t)$. 

SFR is the product of SFE and cold gas mass, i.e.,
${\rm SFR}(t)= {\rm SFE}\cdot M_{\rm gas}(t)$.
According to the expression for the mass of cold gas (see Equation \ref{eq3}), 
the SFR can be written as:
\begin{equation}
{\rm SFR}(t)=\frac{{\rm Const} + \int_{-\infty}^{t} \Phi (t) \cdot {\rm e}^{t/\tau_{\rm dep}} \mathrm{d} t }{\int_{-\infty}^{t}  {\rm e}^{t/\tau_{\rm dep}} \mathrm{d} t}\cdot (1+\lambda)^{-1},
\label{eq5}
\end{equation}
where $\tau_{\rm dep}\propto \rm SFE^{-1} \propto [M_*/R_{\rm e}^2]^{-0.5} $ (see Equation \ref{eq2}) is an effective gas depletion timescale and $\Phi (t)$ is the gas inflow rate on the galaxy scale. $\rm Const$ represents the initial SFR and $\lambda$ is the mass-loading factor of stellar wind. From the above expression, we can clearly see that SFR is determined by the average value of $\Phi (t)$ weighted by ${\rm e}^{t/\tau_{\rm dep}}$ over a timescale of $\tau_{\rm dep}$. This is equivalent to SFR being the result of smoothing $\Phi (t)$ over a time width of $\tau_{\rm dep}$.
The SFR corresponds to the mean gas inflow rate averaged over the gas depletion timescale. Therefore, SFR has smaller fluctuations compared to $\Phi (t)$, and the shorter $\tau_{\rm dep}$ (corresponding larger SFE and $M_*/R_{\rm e}^2$), the larger dispersion of SFR \citep{wang2019, wang2020a, wang2020b}.
Therefore, this explains precisely why the SFR dispersion increases as the radius decreases or the surface density increases, exhibiting symmetry with $\Delta \rm SFMS = 0$ as the axis of symmetry. 

While our model emphasizes the role of gas inflow fluctuations and surface density in setting the SFMS scatter, other physical processes are known to contribute as well. Previous theoretical work has attributed part of the dispersion to stochastic halo accretion \citep{mitra2017}, short-term internal cycles driven by stellar or AGN feedback \citep{matthee2019,katsianis2019}, 
and variations in halo assembly history \citep{tacchella2020}. Hydrodynamical and semi-analytic simulations further show that SNe feedback at low masses and AGN feedback at high masses can enhance the scatter and even reproduce the observed U-shaped $\sigma ({\rm SFR}) - M^*$ relation (e.g.,~\citealt{sparre2017,katsianis2019,legrand2022,huang2023}). Observational evidence for the influence of AGN feedback has also been reported recently \citep{davies2025b}.

\section{The period and amplitude of the accretion flow} 

To determine the period and amplitude of the accretion flow, we quantitatively analyze the relationship between SFR dispersion and stellar surface density for SF galaxies. On the one hand, the connection between $\sigma (\log \rm SFR)$ and the gas inflow rate $\Phi(t)$ is described by Equation \ref{eq4}. On the other hand, the statistical fluctuations of the SFR represent another crucial factor that cannot be overlooked. 
In logarithmic space , the dispersion of the SFR caused by Poisson error can be denoted as:
\begin{equation}
\sigma_{\log \rm SFR}^{\rm stat} \propto \frac{1}{\rm SFR^{1/2}}.
\label{eq6}
\end{equation}
As shown in  Fig. \ref{fig9}, there is a positive correlation between the $\log \rm SFR$ and $\log M_*/R_{\rm e}^2$. Therefore, the SFR dispersion $\sigma (\log \rm SFR)$ caused by statistical fluctuations is inversely correlated with the $\log M_*/R_{\rm e}^2$ (as shown by the blue line in Fig. \ref{fig10}).
In summary, the overall dispersion of the SFR consists of two independent components: the fluctuation of the gas accretion rate and statistical fluctuations. This can be expressed as:
\begin{equation}
\sigma_{\log \rm SFR} ^{\rm total}=\sqrt{{\sigma_{\log \rm SFR}^{\rm \Phi}}^2+{\sigma_{\log \rm SFR}^{\rm stat}}^2}.
\label{eq7}
\end{equation}

In Fig. \ref{fig10}, we fit the observed results using Equation \ref{eq7}, which consists of two components: statistical fluctuations (blue line) and the oscillatory gas inflow rate $\Phi(t)$ (red line). The best-fit model yields an inflow fluctuation period of $T_{\rm P}= 3.50\pm 0.04$ Gyr and an oscillation dispersion of $\sigma_{\Phi}=0.529 \pm 0.003$ dex. These results suggest that cosmic gas accretion flows toward galaxies exhibit a quasi-periodicity of $\sim 3.5$ Gyr, with a characteristic amplitude of 0.5 dex (corresponding to a factor of $\simeq 3$).

We further investigate whether the accretion period correlates with galaxy stellar mass, morphology, or environment. To assess the mass dependence, we divide galaxies into four stellar-mass bins: $\log M_* <9.5$, $9.5<\log M_* <10$, $10<\log M_* <10.5$, and $\log M_* >10.5$. As shown in Fig. \ref{fig11}, the accretion oscillation period decreases systematically with increasing stellar mass, indicating that more massive galaxies undergo faster cycles of inflow fluctuations. This trend can be attributed to the deeper gravitational potentials of massive galaxies, which shorten the dynamical timescales of inflowing gas and thereby accelerate the accretion cycle.
We fit the relation between $T_{\rm P}$ and $M_*$ with a linear model, $\log T_{\rm P}= \alpha \log M_* +\beta$, and use the slope $\alpha$ to quantify the sensitivity of $T_{\rm P}$ to stellar mass. As shown in Fig. \ref{fig11}, the best-fit slopes $\alpha$ for early- and late-type galaxies are $-0.16\pm 0.05$ and $-0.13\pm 0.02$, respectively. This implies that the decline of the accretion period with stellar mass is a bit steeper in early-type galaxies than in late types. A plausible explanation is that the more centrally concentrated gravitational potentials of early-type galaxies render inflow dynamics more sensitive to mass variations, while the extended gas reservoirs and stronger angular-momentum support in late-type galaxies help stabilize inflow, reducing the dependence on mass.
Similarly, the best-fit slopes $\alpha$ for central and satellite galaxies are $-0.05\pm 0.01$ and $-0.15\pm 0.04$, respectively. This indicates that the mass dependence of accretion periods is stronger in satellites than in centrals. A likely explanation is that satellite accretion flows are strongly affected by the tidal field of the host halo and by environmental processes such as ram-pressure stripping and strangulation, which enhance the sensitivity of accretion dynamics to the satellite’s stellar mass. By contrast, central galaxies sustain a more stable, self-regulated inflow within the dominant potential of their host halo.

\begin{figure*}
\centering
\includegraphics[width=1.0\textwidth]{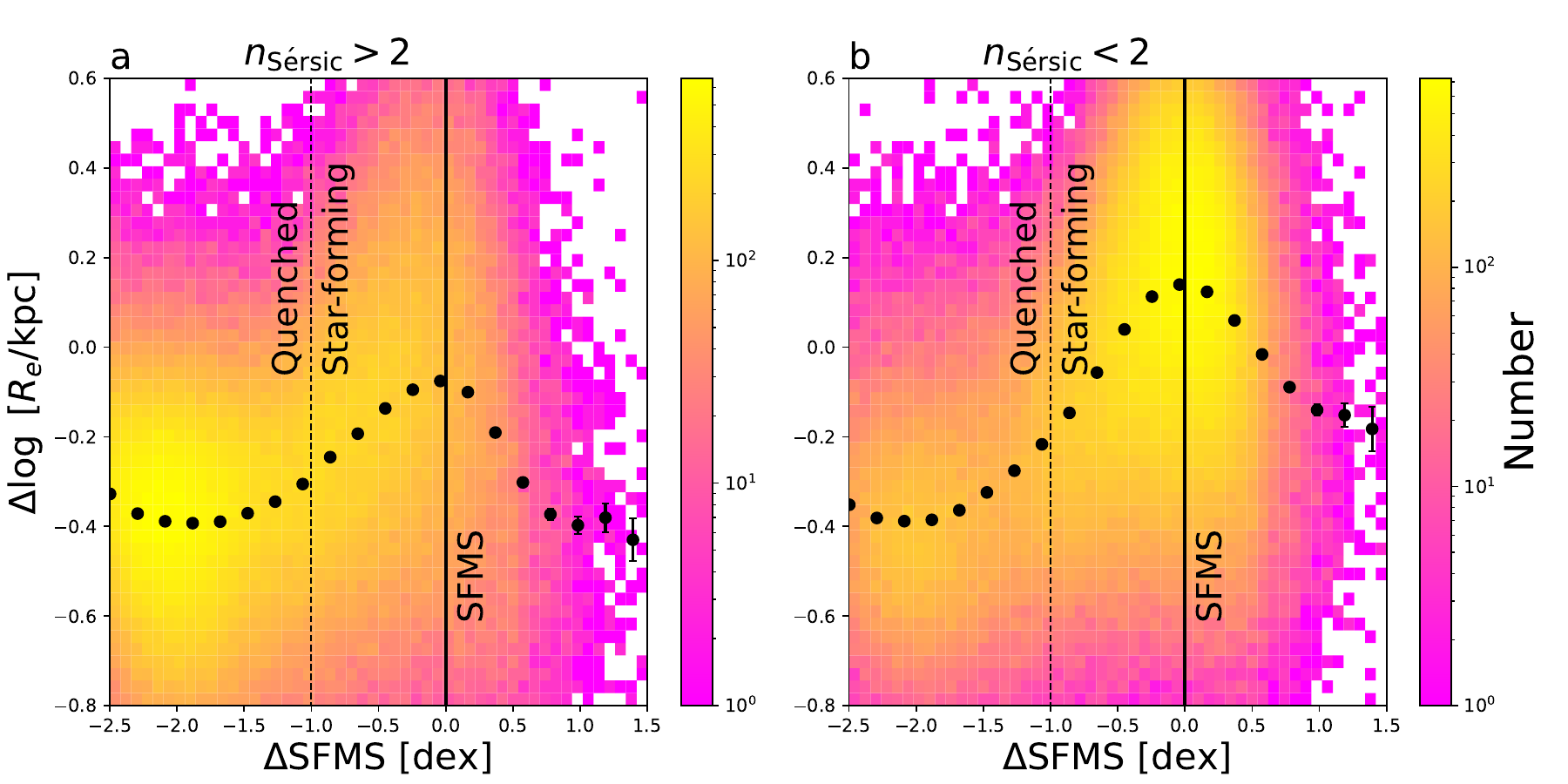}
\caption{\textbf{Comparison of the residual-radius symmetry for subsamples of early-type ($n_{\mathrm{S\acute{e}rsic}}>2$) and late-type galaxies ($n_{\mathrm{S\acute{e}rsic}}<2$).} The residual effective radii, $\Delta \log R_{\rm e}$, display a similar symmetric distribution with respect to the SFMS for both morphological classes, indicating that the symmetry is not driven by galaxy structure. As expected from their more compact stellar configurations, early-type systems show mean $\Delta \log R_{\rm e}$ values that are approximately 0.2 dex smaller than those of late-type disk galaxies. Nevertheless, the overall symmetric pattern is preserved across both subsamples, highlighting the robustness of the radius symmetry along the SFMS.
}\label{fig12}
\end{figure*}

\begin{figure*}
\centering
\includegraphics[width=1.0\textwidth]{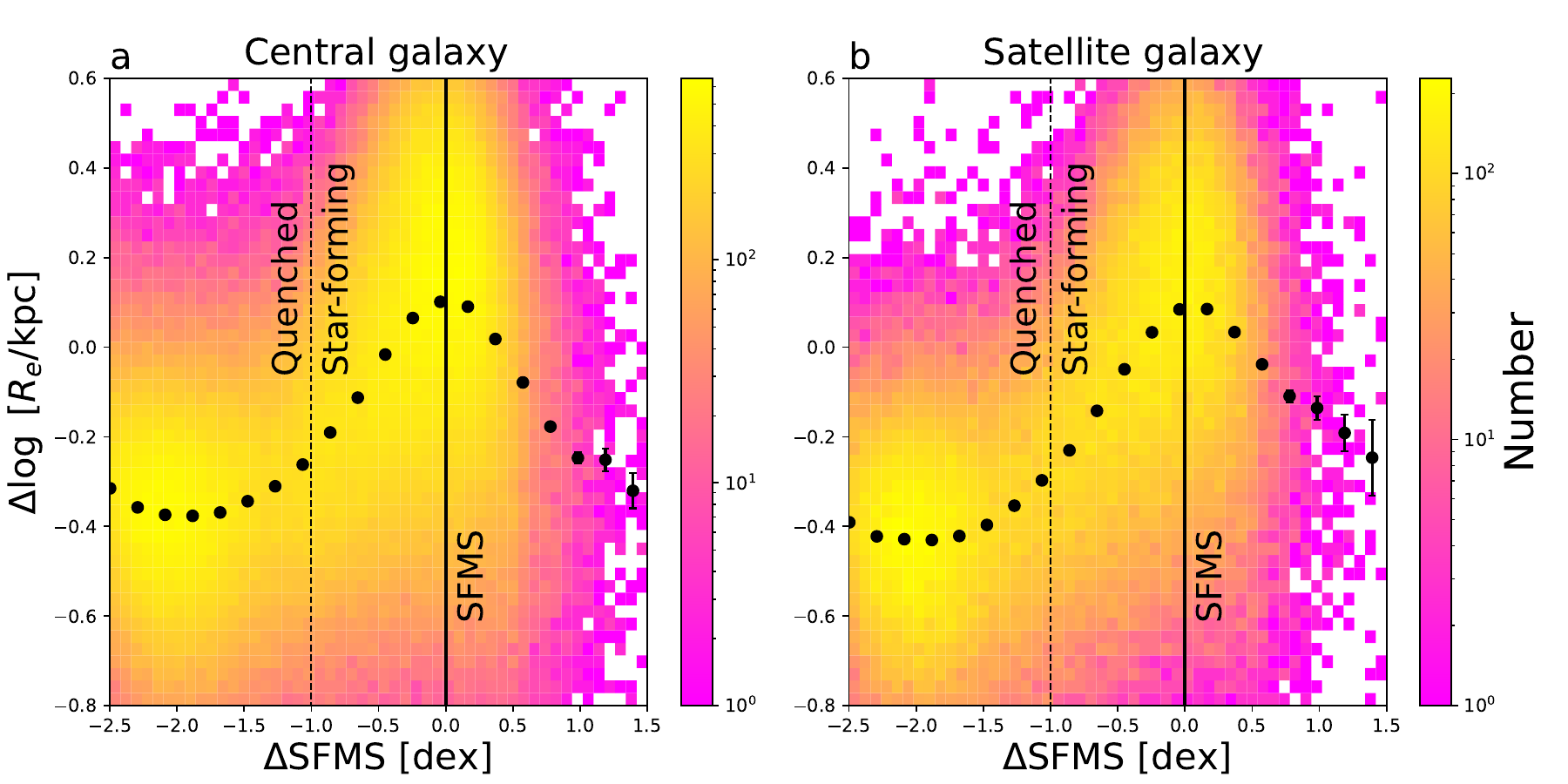}
\caption{\textbf{Comparison of the residual-radius symmetry for subsamples of central and satellite galaxies.} The residual effective radii, $\Delta \log R_{\rm e}$, show comparably symmetric distributions with respect to the SFMS for both central and satellite systems. Despite their differing environmental histories, the two populations display nearly identical symmetry patterns, indicating that the radius symmetry along the SFMS is largely insensitive to whether a galaxy resides at the center of its halo or within a satellite configuration.
}\label{fig13}
\end{figure*}

\section{Universality of Radius Symmetry Along the SFMS}

Here we investigate whether the symmetry of effective radii along the SFMS represents a universal property of SF galaxies or whether it is confined to particular morphological or environmental classes. As shown in Fig. \ref{fig12}, both early-type systems ($n_{\rm S\acute{e}rsic}>2$) and late-type systems ($n_{\rm S\acute{e}rsic}<2$) exhibit a clear symmetry in their residual effective radius, $\Delta\log R_{\rm e}$, with respect to the SFMS. As expected, early-type galaxies are systematically more compact, with their mean $\Delta\log R_{\rm e}$ smaller by $\sim$0.2 dex compared to late-type galaxies, yet the symmetric pattern persists across both classes. Likewise, Fig. \ref{fig13} shows that central and satellite galaxies display the same symmetric distribution in $\Delta\log R_{\rm e}$ despite their distinct environmental conditions.

Taken together, these results indicate that the radius symmetry along the SFMS is not an artifact of morphology or environment but instead reflects a fundamental and robust structural property of star-forming galaxies. The persistence of this symmetry across early vs. late types and central vs. satellite systems suggests that it is likely governed by a common underlying physical mechanism, potentially linked to the self-regulated nature of star formation and the coupled evolution of galaxy size and stellar mass.

\section{Conclusion}

In this work, we investigate the primary factors shaping the dispersion of the SFMS. Using a large galaxy sample from SDSS DR7, we find that the intrinsic scatter of the SFMS is mainly regulated by stellar mass and effective radius. In particular, both $R_{\rm e}$ and $M_*/R_{\rm e}^2$ exhibit a remarkable symmetry with respect to the SFMS ridgeline, with deviations systematically associated with smaller radii or larger surface stellar mass densities.
By combining these structural trends with the continuity equation, we infer that the SFMS dispersion originates from fluctuations in galaxy gas accretion flows, while the SFR traces the time-averaged accretion rate over the gas depletion timescale. Galaxies with shorter depletion timescales—typically those with higher $M_*/R_{\rm e}^2$—naturally experience larger fluctuations around the SFMS. Additional statistical fluctuations further broaden the observed dispersion, and the inclusion of quenched systems enhances both the scatter and the asymmetry of the sequence.

Taken together, our results demonstrate that the symmetry of galaxy radii and surface stellar mass density along the SFMS is a robust and universal feature—present across different structural, morphological, and environmental populations. These findings provide new insights into the interplay among cosmic gas accretion, internal galaxy structure, and star-formation activity, and they advance our understanding of the physical origins of the SFMS and its dispersion.

\section{Acknowledgements}
We appreciate the reviewer’s thorough review and constructive comments, which have been very helpful in strengthening the manuscript.
This work was supported by the National SKA Program of China (No. 2025SKA0130100). Z. C. He is supported by the USTC Research Launch Project KY2030000187 and the National Natural Science Foundation of China (Grant Nos. 12222304, 12192220, and 12192221). E. Wang is supported by the National Natural Science Foundation of China, Grant Nos. 12473008. H.Y. W. is supported by the National Natural Science Foundation of China Nos. 12192224 and CAS Project for Young Scientists in Basic Research, Grant No. YSBR-062.

\section{Data Availability Statement}
The data underlying this article will be shared on reasonable request to the corresponding author.

\bibliographystyle{mnras}

\begin{thebibliography}{}
\makeatletter
\relax
\def\mn@urlcharsother{\let\do\@makeother \do\$\do\&\do\#\do\^\do\_\do\%\do\~}
\def\mn@doi{\begingroup\mn@urlcharsother \@ifnextchar [ {\mn@doi@}
  {\mn@doi@[]}}
\def\mn@doi@[#1]#2{\def\@tempa{#1}\ifx\@tempa\@empty \href
  {http://dx.doi.org/#2} {doi:#2}\else \href {http://dx.doi.org/#2} {#1}\fi
  \endgroup}
\def\mn@eprint#1#2{\mn@eprint@#1:#2::\@nil}
\def\mn@eprint@arXiv#1{\href {http://arxiv.org/abs/#1} {{\tt arXiv:#1}}}
\def\mn@eprint@dblp#1{\href {http://dblp.uni-trier.de/rec/bibtex/#1.xml}
  {dblp:#1}}
\def\mn@eprint@#1:#2:#3:#4\@nil{\def\@tempa {#1}\def\@tempb {#2}\def\@tempc
  {#3}\ifx \@tempc \@empty \let \@tempc \@tempb \let \@tempb \@tempa \fi \ifx
  \@tempb \@empty \def\@tempb {arXiv}\fi \@ifundefined
  {mn@eprint@\@tempb}{\@tempb:\@tempc}{\expandafter \expandafter \csname
  mn@eprint@\@tempb\endcsname \expandafter{\@tempc}}}

\bibitem[\protect\citeauthoryear{Ayromlou, Kauffmann, Yates, Nelson  \&
  White}{Ayromlou et~al.}{2021}]{ayromlou2021}
Ayromlou M.,  Kauffmann G.,  Yates R.~M.,  Nelson D.,   White S.~D.,  2021,
  Monthly Notices of the Royal Astronomical Society, 505, 492

\bibitem[\protect\citeauthoryear{Boogaard et~al.,}{Boogaard
  et~al.}{2018}]{boogaard2018}
Boogaard L.~A.,  et~al., 2018, Astronomy \& Astrophysics, 619, A27

\bibitem[\protect\citeauthoryear{Brinchmann, Charlot, White, Tremonti,
  Kauffmann, Heckman  \& Brinkmann}{Brinchmann et~al.}{2004}]{brinchmann2004}
Brinchmann J.,  Charlot S.,  White S.~D.,  Tremonti C.,  Kauffmann G.,  Heckman
  T.,   Brinkmann J.,  2004, Monthly notices of the royal astronomical society,
  351, 1151

\bibitem[\protect\citeauthoryear{{Caplar} \& {Tacchella}}{{Caplar} \&
  {Tacchella}}{2019}]{Caplar-19}
{Caplar} N.,  {Tacchella} S.,  2019, \mn@doi [\mnras] {10.1093/mnras/stz1449},
  \href {https://ui.adsabs.harvard.edu/abs/2019MNRAS.487.3845C} {487, 3845}

\bibitem[\protect\citeauthoryear{Chen et~al.,}{Chen et~al.}{2012}]{chen2012}
Chen Y.-M.,  et~al., 2012, Monthly Notices of the Royal Astronomical Society,
  421, 314

\bibitem[\protect\citeauthoryear{Chen, He, Ho, Gu, Wang, Zhuang, Liu  \&
  Wang}{Chen et~al.}{2022}]{chen2022}
Chen Z.,  He Z.,  Ho L.~C.,  Gu Q.,  Wang T.,  Zhuang M.,  Liu G.,   Wang Z.,
  2022, Nature Astronomy, 6, 339

\bibitem[\protect\citeauthoryear{Cicone et~al.,}{Cicone
  et~al.}{2014}]{cicone2014}
Cicone C.,  et~al., 2014, Astronomy \& Astrophysics, 562, A21

\bibitem[\protect\citeauthoryear{Citro, Pozzetti, Moresco  \& Cimatti}{Citro
  et~al.}{2016}]{citro2016}
Citro A.,  Pozzetti L.,  Moresco M.,   Cimatti A.,  2016, Astronomy \&
  Astrophysics, 592, A19

\bibitem[\protect\citeauthoryear{Clarke, Shapley, Sanders, Topping, Brammer,
  Bento, Reddy  \& Kehoe}{Clarke et~al.}{2024}]{clarke2024}
Clarke L.,  Shapley A.~E.,  Sanders R.~L.,  Topping M.~W.,  Brammer G.~B.,
  Bento T.,  Reddy N.~A.,   Kehoe E.,  2024, The Astrophysical Journal, 977,
  133

\bibitem[\protect\citeauthoryear{Daddi et~al.,}{Daddi et~al.}{2007}]{daddi2007}
Daddi E.,  et~al., 2007, The Astrophysical Journal, 670, 156

\bibitem[\protect\citeauthoryear{Davies et~al.,}{Davies
  et~al.}{2019}]{davies2019}
Davies L.,  et~al., 2019, Monthly Notices of the Royal Astronomical Society,
  483, 1881

\bibitem[\protect\citeauthoryear{Davies et~al.,}{Davies
  et~al.}{2025a}]{davies2025a}
Davies L.,  et~al., 2025a, Monthly Notices of the Royal Astronomical Society,
  540, 1870

\bibitem[\protect\citeauthoryear{Davies et~al.,}{Davies
  et~al.}{2025b}]{davies2025b}
Davies L.,  et~al., 2025b, Monthly Notices of the Royal Astronomical Society,
  541, 573

\bibitem[\protect\citeauthoryear{Donnari, Pillepich, Nelson, Marinacci,
  Vogelsberger  \& Hernquist}{Donnari et~al.}{2021}]{donnari2021}
Donnari M.,  Pillepich A.,  Nelson D.,  Marinacci F.,  Vogelsberger M.,
  Hernquist L.,  2021, Monthly Notices of the Royal Astronomical Society, 506,
  4760

\bibitem[\protect\citeauthoryear{{Dopita} \& {Ryder}}{{Dopita} \&
  {Ryder}}{1994}]{Dopita1994}
{Dopita} M.~A.,  {Ryder} S.~D.,  1994, \mn@doi [\apj] {10.1086/174390}, \href
  {https://ui.adsabs.harvard.edu/abs/1994ApJ...430..163D} {430, 163}

\bibitem[\protect\citeauthoryear{Elbaz et~al.,}{Elbaz et~al.}{2007}]{elbaz2007}
Elbaz D.,  et~al., 2007, Astronomy \& Astrophysics, 468, 33

\bibitem[\protect\citeauthoryear{Elbaz et~al.,}{Elbaz et~al.}{2011}]{elbaz2011}
Elbaz D.,  et~al., 2011, Astronomy \& Astrophysics, 533, A119

\bibitem[\protect\citeauthoryear{Fabian}{Fabian}{2012}]{fabian2012}
Fabian A.~C.,  2012, Annual Review of Astronomy and Astrophysics, 50, 455

\bibitem[\protect\citeauthoryear{Franx, van Dokkum, Schreiber, Wuyts, Labb{\'e}
   \& Toft}{Franx et~al.}{2008}]{franx2008}
Franx M.,  van Dokkum P.~G.,  Schreiber N. M.~F.,  Wuyts S.,  Labb{\'e} I.,
  Toft S.,  2008, The Astrophysical Journal, 688, 770

\bibitem[\protect\citeauthoryear{Genzel et~al.,}{Genzel
  et~al.}{2015}]{genzel2015}
Genzel R.,  et~al., 2015, The Astrophysical Journal, 800, 20

\bibitem[\protect\citeauthoryear{Guo, Zheng, Wang  \& Fu}{Guo
  et~al.}{2015}]{guo2015}
Guo K.,  Zheng X.~Z.,  Wang T.,   Fu H.,  2015, The Astrophysical Journal
  Letters, 808, L49

\bibitem[\protect\citeauthoryear{He et~al.,}{He et~al.}{2019}]{he2019}
He Z.,  et~al., 2019, Nature Astronomy, 3, 265

\bibitem[\protect\citeauthoryear{He et~al.,}{He et~al.}{2022}]{he2022}
He Z.,  et~al., 2022, Science Advances, 8, eabk3291

\bibitem[\protect\citeauthoryear{He et~al.,}{He et~al.}{2024}]{he2024}
He Z.,  et~al., 2024, Science China Physics, Mechanics \& Astronomy, 67, 129512

\bibitem[\protect\citeauthoryear{Huang, Battisti, Grasha, da Cunha, Lagos,
  Leslie  \& Wisnioski}{Huang et~al.}{2023}]{huang2023}
Huang R.,  Battisti A.~J.,  Grasha K.,  da Cunha E.,  Lagos C. d.~P.,  Leslie
  S.~K.,   Wisnioski E.,  2023, Monthly Notices of the Royal Astronomical
  Society, 520, 446

\bibitem[\protect\citeauthoryear{Ishibashi \& Fabian}{Ishibashi \&
  Fabian}{2012}]{ishibashi2012}
Ishibashi W.,  Fabian A.,  2012, Monthly Notices of the Royal Astronomical
  Society, 427, 2998

\bibitem[\protect\citeauthoryear{Katsianis et~al.,}{Katsianis
  et~al.}{2019}]{katsianis2019}
Katsianis A.,  et~al., 2019, The Astrophysical Journal, 879, 11

\bibitem[\protect\citeauthoryear{King \& Pounds}{King \&
  Pounds}{2015}]{king2015}
King A.,  Pounds K.,  2015, Annual Review of Astronomy and Astrophysics, 53,
  115

\bibitem[\protect\citeauthoryear{Kurczynski et~al.,}{Kurczynski
  et~al.}{2016}]{kurczynski2016}
Kurczynski P.,  et~al., 2016, The Astrophysical Journal Letters, 820, L1

\bibitem[\protect\citeauthoryear{Legrand, Hutter, Dayal, Ucci, Gottl{\"o}ber
  \& Yepes}{Legrand et~al.}{2022}]{legrand2022}
Legrand L.,  Hutter A.,  Dayal P.,  Ucci G.,  Gottl{\"o}ber S.,   Yepes G.,
  2022, Monthly Notices of the Royal Astronomical Society, 509, 595

\bibitem[\protect\citeauthoryear{Lilly, Carollo, Pipino, Renzini  \&
  Peng}{Lilly et~al.}{2013}]{lilly2013}
Lilly S.~J.,  Carollo C.~M.,  Pipino A.,  Renzini A.,   Peng Y.,  2013, The
  Astrophysical Journal, 772, 119

\bibitem[\protect\citeauthoryear{Matthee \& Schaye}{Matthee \&
  Schaye}{2019}]{matthee2019}
Matthee J.,  Schaye J.,  2019, Monthly Notices of the Royal Astronomical
  Society, 484, 915

\bibitem[\protect\citeauthoryear{Meert, Vikram  \& Bernardi}{Meert
  et~al.}{2015}]{meert2015}
Meert A.,  Vikram V.,   Bernardi M.,  2015, Monthly Notices of the Royal
  Astronomical Society, 446, 3943

\bibitem[\protect\citeauthoryear{Meert, Vikram  \& Bernardi}{Meert
  et~al.}{2016}]{meert2016}
Meert A.,  Vikram V.,   Bernardi M.,  2016, Monthly Notices of the Royal
  Astronomical Society, 455, 2440

\bibitem[\protect\citeauthoryear{Mendel, Simard, Palmer, Ellison  \&
  Patton}{Mendel et~al.}{2013}]{mendel2013}
Mendel J.~T.,  Simard L.,  Palmer M.,  Ellison S.~L.,   Patton D.~R.,  2013,
  The Astrophysical Journal Supplement Series, 210, 3

\bibitem[\protect\citeauthoryear{Mitra, Dav{\'e}, Simha  \& Finlator}{Mitra
  et~al.}{2017}]{mitra2017}
Mitra S.,  Dav{\'e} R.,  Simha V.,   Finlator K.,  2017, Monthly Notices of the
  Royal Astronomical Society, 464, 2766

\bibitem[\protect\citeauthoryear{Morselli, Popesso, Cibinel, Oesch, Montes,
  Atek, Illingworth  \& Holden}{Morselli et~al.}{2019}]{morselli2019}
Morselli L.,  Popesso P.,  Cibinel A.,  Oesch P.~A.,  Montes M.,  Atek H.,
  Illingworth G.~D.,   Holden B.,  2019, Astronomy \& Astrophysics, 626, A61

\bibitem[\protect\citeauthoryear{Noeske et~al.,}{Noeske
  et~al.}{2007}]{noeske2007}
Noeske K.,  et~al., 2007, The Astrophysical Journal, 660, L43

\bibitem[\protect\citeauthoryear{Old et~al.,}{Old et~al.}{2020}]{old2020}
Old L.~J.,  et~al., 2020, Monthly Notices of the Royal Astronomical Society,
  493, 5987

\bibitem[\protect\citeauthoryear{Pedregosa et~al.,}{Pedregosa
  et~al.}{2011}]{pedregosa2011}
Pedregosa F.,  et~al., 2011, the Journal of machine Learning research, 12, 2825

\bibitem[\protect\citeauthoryear{Peng et~al.,}{Peng et~al.}{2010}]{peng2010}
Peng Y.-j.,  et~al., 2010, The Astrophysical Journal, 721, 193

\bibitem[\protect\citeauthoryear{Popesso et~al.,}{Popesso
  et~al.}{2023}]{popesso2023}
Popesso P.,  et~al., 2023, Monthly Notices of the Royal Astronomical Society,
  519, 1526

\bibitem[\protect\citeauthoryear{Quai, Pozzetti, Citro, Moresco  \&
  Cimatti}{Quai et~al.}{2018}]{quai2018}
Quai S.,  Pozzetti L.,  Citro A.,  Moresco M.,   Cimatti A.,  2018, Monthly
  Notices of the Royal Astronomical Society, 478, 3335

\bibitem[\protect\citeauthoryear{Rasmussen, Mulchaey, Bai, Ponman, Raychaudhury
   \& Dariush}{Rasmussen et~al.}{2012}]{rasmussen2012}
Rasmussen J.,  Mulchaey J.~S.,  Bai L.,  Ponman T.~J.,  Raychaudhury S.,
  Dariush A.,  2012, The Astrophysical Journal, 757, 122

\bibitem[\protect\citeauthoryear{Renzini \& Peng}{Renzini \&
  Peng}{2015}]{renzini2015}
Renzini A.,  Peng Y.-j.,  2015, The Astrophysical Journal Letters, 801, L29

\bibitem[\protect\citeauthoryear{Rodighiero et~al.,}{Rodighiero
  et~al.}{2010}]{rodighiero2010}
Rodighiero G.,  et~al., 2010, Astronomy \& Astrophysics, 518, L25

\bibitem[\protect\citeauthoryear{Salim et~al.,}{Salim et~al.}{2007}]{salim2007}
Salim S.,  et~al., 2007, The Astrophysical Journal Supplement Series, 173, 267

\bibitem[\protect\citeauthoryear{Salim et~al.,}{Salim et~al.}{2016}]{salim2016}
Salim S.,  et~al., 2016, The Astrophysical Journal Supplement Series, 227, 2

\bibitem[\protect\citeauthoryear{Salim, Boquien  \& Lee}{Salim
  et~al.}{2018}]{salim2018}
Salim S.,  Boquien M.,   Lee J.~C.,  2018, The Astrophysical Journal, 859, 11

\bibitem[\protect\citeauthoryear{Santini et~al.,}{Santini
  et~al.}{2017}]{santini2017}
Santini P.,  et~al., 2017, The Astrophysical Journal, 847, 76

\bibitem[\protect\citeauthoryear{Schiminovich et~al.,}{Schiminovich
  et~al.}{2007}]{schiminovich2007}
Schiminovich D.,  et~al., 2007, The Astrophysical Journal Supplement Series,
  173, 315

\bibitem[\protect\citeauthoryear{Schreiber et~al.,}{Schreiber
  et~al.}{2015}]{schreiber2015}
Schreiber C.,  et~al., 2015, Astronomy \& Astrophysics, 575, A74

\bibitem[\protect\citeauthoryear{Shi, Helou, Yan, Armus, Wu, Papovich  \&
  Stierwalt}{Shi et~al.}{2011}]{shi2011}
Shi Y.,  Helou G.,  Yan L.,  Armus L.,  Wu Y.,  Papovich C.,   Stierwalt S.,
  2011, The Astrophysical Journal, 733, 87

\bibitem[\protect\citeauthoryear{Shi et~al.,}{Shi et~al.}{2018}]{shi2018}
Shi Y.,  et~al., 2018, The Astrophysical Journal, 853, 149

\bibitem[\protect\citeauthoryear{Sparre, Hayward, Feldmann,
  Faucher-Gigu{\`e}re, Muratov, Kere{\v{s}}  \& Hopkins}{Sparre
  et~al.}{2017}]{sparre2017}
Sparre M.,  Hayward C.~C.,  Feldmann R.,  Faucher-Gigu{\`e}re C.-A.,  Muratov
  A.~L.,  Kere{\v{s}} D.,   Hopkins P.~F.,  2017, Monthly Notices of the Royal
  Astronomical Society, 466, 88

\bibitem[\protect\citeauthoryear{Speagle, Steinhardt, Capak  \&
  Silverman}{Speagle et~al.}{2014}]{speagle2014}
Speagle J.~S.,  Steinhardt C.~L.,  Capak P.~L.,   Silverman J.~D.,  2014, The
  Astrophysical Journal Supplement Series, 214, 15

\bibitem[\protect\citeauthoryear{Tacchella et~al.,}{Tacchella
  et~al.}{2015}]{tacchella2015}
Tacchella S.,  et~al., 2015, Science, 348, 314

\bibitem[\protect\citeauthoryear{Tacchella, Forbes  \& Caplar}{Tacchella
  et~al.}{2020}]{tacchella2020}
Tacchella S.,  Forbes J.~C.,   Caplar N.,  2020, Monthly Notices of the Royal
  Astronomical Society, 497, 698

\bibitem[\protect\citeauthoryear{Tacconi et~al.,}{Tacconi
  et~al.}{2018}]{tacconi2018}
Tacconi L.~J.,  et~al., 2018, The Astrophysical Journal, 853, 179

\bibitem[\protect\citeauthoryear{Tamburri, Saracco, Longhetti, Gargiulo, Lonoce
   \& Ciocca}{Tamburri et~al.}{2014}]{tamburri2014}
Tamburri S.,  Saracco P.,  Longhetti M.,  Gargiulo A.,  Lonoce I.,   Ciocca F.,
   2014, Astronomy \& Astrophysics, 570, A102

\bibitem[\protect\citeauthoryear{Thorne et~al.,}{Thorne
  et~al.}{2021}]{thorne2021}
Thorne J.~E.,  et~al., 2021, Monthly Notices of the Royal Astronomical Society,
  505, 540

\bibitem[\protect\citeauthoryear{Tomczak et~al.,}{Tomczak
  et~al.}{2016}]{tomczak2016}
Tomczak A.~R.,  et~al., 2016, The Astrophysical Journal, 817, 118

\bibitem[\protect\citeauthoryear{Veilleux, Maiolino, Bolatto  \&
  Aalto}{Veilleux et~al.}{2020}]{veilleux2020}
Veilleux S.,  Maiolino R.,  Bolatto A.~D.,   Aalto S.,  2020, The Astronomy and
  Astrophysics Review, 28, 1

\bibitem[\protect\citeauthoryear{{Wang} \& {Lilly}}{{Wang} \&
  {Lilly}}{2020a}]{wang2020a}
{Wang} E.,  {Lilly} S.~J.,  2020a, \mn@doi [\apj] {10.3847/1538-4357/ab7b7d},
  \href {https://ui.adsabs.harvard.edu/abs/2020ApJ...892...87W} {892, 87}

\bibitem[\protect\citeauthoryear{{Wang} \& {Lilly}}{{Wang} \&
  {Lilly}}{2020b}]{wang2020b}
{Wang} E.,  {Lilly} S.~J.,  2020b, \mn@doi [\apj] {10.3847/1538-4357/ab8b5e},
  \href {https://ui.adsabs.harvard.edu/abs/2020ApJ...895...25W} {895, 25}

\bibitem[\protect\citeauthoryear{Wang, Kong  \& Pan}{Wang
  et~al.}{2018}]{wang2018}
Wang E.,  Kong X.,   Pan Z.,  2018, The Astrophysical Journal, 865, 49

\bibitem[\protect\citeauthoryear{Wang, Lilly, Pezzulli  \& Matthee}{Wang
  et~al.}{2019}]{wang2019}
Wang E.,  Lilly S.~J.,  Pezzulli G.,   Matthee J.,  2019, The Astrophysical
  Journal, 877, 132

\bibitem[\protect\citeauthoryear{Wetzel, Tinker  \& Conroy}{Wetzel
  et~al.}{2012}]{wetzel2012}
Wetzel A.~R.,  Tinker J.~L.,   Conroy C.,  2012, Monthly Notices of the Royal
  Astronomical Society, 424, 232

\bibitem[\protect\citeauthoryear{Whitaker, Van~Dokkum, Brammer  \&
  Franx}{Whitaker et~al.}{2012}]{whitaker2012}
Whitaker K.~E.,  Van~Dokkum P.~G.,  Brammer G.,   Franx M.,  2012, The
  Astrophysical Journal Letters, 754, L29

\bibitem[\protect\citeauthoryear{Whitaker et~al.,}{Whitaker
  et~al.}{2014}]{whitaker2014}
Whitaker K.~E.,  et~al., 2014, The Astrophysical Journal, 795, 104

\bibitem[\protect\citeauthoryear{Wuyts et~al.,}{Wuyts et~al.}{2011}]{wuyts2011}
Wuyts S.,  et~al., 2011, The Astrophysical Journal, 742, 96

\bibitem[\protect\citeauthoryear{Yang, Mo, Van~den Bosch, Pasquali, Li  \&
  Barden}{Yang et~al.}{2007}]{yang2007}
Yang X.,  Mo H.,  Van~den Bosch F.~C.,  Pasquali A.,  Li C.,   Barden M.,
  2007, The Astrophysical Journal, 671, 153

\bibitem[\protect\citeauthoryear{{Yesuf}, {Ho}  \& {Faber}}{{Yesuf}
  et~al.}{2021}]{Yesuf2021}
{Yesuf} H.~M.,  {Ho} L.~C.,   {Faber} S.~M.,  2021, \mn@doi [\apj]
  {10.3847/1538-4357/ac27a7}, \href
  {https://ui.adsabs.harvard.edu/abs/2021ApJ...923..205Y} {923, 205}

\bibitem[\protect\citeauthoryear{{Yu}, {Ho}  \& {Wang}}{{Yu}
  et~al.}{2021}]{Yu2021}
{Yu} S.-Y.,  {Ho} L.~C.,   {Wang} J.,  2021, \mn@doi [\apj]
  {10.3847/1538-4357/ac0c77}, \href
  {https://ui.adsabs.harvard.edu/abs/2021ApJ...917...88Y} {917, 88}

\bibitem[\protect\citeauthoryear{{Yu}, {Xu}, {Ho}, {Wang}  \& {Kao}}{{Yu}
  et~al.}{2022a}]{Yu2022}
{Yu} S.-Y.,  {Xu} D.,  {Ho} L.~C.,  {Wang} J.,   {Kao} W.-B.,  2022a, \mn@doi
  [\aap] {10.1051/0004-6361/202142533}, \href
  {https://ui.adsabs.harvard.edu/abs/2022A&A...661A..98Y} {661, A98}

\bibitem[\protect\citeauthoryear{{Yu} et~al.,}{{Yu} et~al.}{2022b}]{Yu2022b}
{Yu} S.-Y.,  et~al., 2022b, \mn@doi [\aap] {10.1051/0004-6361/202244306}, \href
  {https://ui.adsabs.harvard.edu/abs/2022A&A...666A.175Y} {666, A175}

\bibitem[\protect\citeauthoryear{Zubovas \& King}{Zubovas \&
  King}{2012}]{zubovas2012}
Zubovas K.,  King A.,  2012, The Astrophysical Journal Letters, 745, L34

\makeatother
\end{thebibliography}

\clearpage



\end{document}